\documentclass[prd,twocolumn,showpacs,preprintnumbers,unsortedaddress,a4paper,superscriptaddress]{revtex4}

\usepackage{amsfonts}
\usepackage{amsmath}
\usepackage{amssymb}
\usepackage{aas_macros}
\usepackage{graphicx}
\usepackage{color}
\usepackage{hyperref}
\usepackage{ulem}

\allowdisplaybreaks

\definecolor{RoyalBlue}{rgb}{0.25,.41,.88}

\def\lsim{~\rlap{$<$}{\lower 1.0ex\hbox{$\sim$}}}
\def\bsim{~\rlap{$>$}{\lower 1.0ex\hbox{$\sim$}}}

\def\tr{{\rm tr}}
\def\det{{\rm det}}

\newcommand{\bm}[1]{{\mbox{\boldmath $#1$}}}

\def\veta{{\boldsymbol\eta}}

\def\vomg{{\boldsymbol\Omega}}

\def\vk{\mathrm{\bf k}}
\def\vn{\mathrm{\bf n}}
\def\vm{\mathrm{\bf m}}

\def\vw{\mathrm{\bf w}}
\def\vx{\mathrm{\bf x}}
\def\vy{\mathrm{\bf y}}

\def\vii{\mathrm{I}}

\def\grad{\bm{\nabla}}

\def\fnl{f_{\rm NL}}
\def\gnl{g_{\rm NL}}

\def\npk{n_{\rm pk}}

\def\cpk{n_\text{pk}}
\def\bnpk{\bar{n}_{\rm pk}}

\def\dpk{\delta_{\rm pk}}

\def\xpk{\xi_{\rm pk}}

\def\apjl{Astrophys.\ J.\ Lett.}

\def\aap{Astron.\ Astrophys.}

\def\apjs{Astrophys.\ J. Supp.}

\newcommand{\mb}{\mathbf}

\newcommand{\avg}[1]{\left\langle #1 \right\rangle}
\newcommand{\pd}{\partial}


\begin{document}

\bibliographystyle{revtex}

\title[Lagrangian bias of generic large-scale structure tracers]{Lagrangian bias of generic LSS tracers}

\author{Titouan Lazeyras} \email{titouan@mpa-garching.mpg.de}
\affiliation{Max-Planck-Institut f\"ur Astrophysik, Karl-Schwarzschild-Str. 1, 
85748 Garching b. M\"unchen, Germany}

\author{Marcello Musso} \email{mmusso@sas.upenn.edu}
\affiliation{Max-Planck-Institut f\"ur Astrophysik, Karl-Schwarzschild-Str. 1, 
85748 Garching b. M\"unchen, Germany}
\affiliation{Center for Particle Cosmology, University of Pennsylvania, 209 S 33rd Street, 
  Philadelphia, PA 19104, USA}

\author{Vincent Desjacques} \email{Vincent.Desjacques@unige.ch}
\affiliation{D\'epartement de Physique Th\'eorique and Center for Astroparticle Physics (CAP)
Universit\'e de Gen\`eve, 24 quai Ernest Ansermet, CH-1211 Gen\`eve, Switzerland}

\begin{abstract}

The dark matter halos that host galaxies and clusters form out of initial high-density patches, providing a biased tracer of the linear matter density field. 
In the simplest local bias approximation, the halo field is treated as a perturbative series in the average overdensity of the Lagrangian patch.
In more realistic models, however, additional quantities will affect the clustering of halo-patches, and this expansion becomes a function of several stochastic variables. 
In this paper, we present a general multivariate expansion scheme that can parametrize the clustering of any biased Lagrangian tracer, given only the variables involved and their symmetry (in our case rotational invariance). This approach is based on an expansion in the orthonormal polynomials associated with the relevant variables, so that no renormalization of the coefficients ever occurs.
We provide explicit expression for the series coefficients, or Lagrangian bias parameters, in the case of peaks of the linear density field. As an application of our formalism, we present a simple derivation 
of the original BBKS formula, and compute the non-Gaussian bias in the presence of a primordial trispectrum of the local shape.

\end{abstract}

\maketitle

\section{Introduction}
\label{sec:intro}

Since the original work of \cite{kaiser:1984}, perturbative bias expansions have been extensively 
used to model the clustering of biased tracers of the large scale structure down to mildly nonlinear 
scales (see, for instance,
\cite{politzer/wise:1984,jensen/szalay:1986,szalay:1988,fry/gaztanaga:1993,matarrese/verde/heavens:1997,
taruya/soda:1999,catelan/porciani/kamionkowski:2000,gaztanaga/scoccimarro:2005,matsubara:2008,
mcdonald/roy:2009,giannantonio/porciani:2010,schmidt/jeong/desjacques:2013,sheth/chan/scoccimarro:2013}). 
However, many of these bias expansions are not well behaved as they generate corrections at all orders 
which strongly depend on the amount of smoothing applied to the density field 
\cite{heavens/matarrese/verde:1998}. 
To resolve this issue, ref.~\cite{mcdonald:2006} suggested to redefine the free parameters of the model 
(i.e. the coefficients of the original Taylor expansion) so that they closely correspond to observables 
and, therefore, are independent of the smoothing introduced to justify the Taylor expansion. 
This approach, which has been applied in e.g. 
\cite{jeong/komatsu:2009,saito/takada/taruya:2009,matsubara:2011}, has the advantage that it can been
carried out both in Lagrangian and Eulerian space.

When biasing is defined in Lagrangian space, series expansions in orthogonal polynomials provide an 
alternative solution to this problem. Such expansions have already been considered in the literature, 
e.g. in \cite[][]{szalay:1988,coles:1993,matsubara:1995,ferraro/smith/etal:2013}.
Ref.~\cite{szalay:1988} (hereafter S88) adopted a simple sharp clipping (i.e. a threshold on the value) 
of the linear density field as a 
proxy for biasing, which can be expanded in a basis of univariate Hermite polynomials. 
One can then easily demonstrate that the coefficients of this expansion are the renormalized bias parameters, 
and recover some of the results of \cite{schmidt/jeong/desjacques:2013}. 
Furthermore, S88 also pointed out that the same method can be 
applied to any biasing scheme defined as a set of constraints in Lagrangian space, leading to an expansion 
in a series of multivariate Hermite polynomials.
Unfortunately, calculations with multivariate Hermite polynomials are not very illuminating, and rapidly 
become tedious when there are several variables like in a peak constraint \cite{bardeen/bond/etal:1986} 
(hereafter BBKS). 
In this case, exploiting the invariance under rotations as done in \cite{desjacques:2013} (hereafter D13) 
provides a much more efficient and informative way of writing down the perturbative bias expansion.

In this paper, we will extend the work of S88 and D13 and derive a generic Lagrangian perturbative bias 
expansion compatible with rotational invariance. Our approach can be applied to any tracer of the large 
scale structure that can be described by an arbitrary set of constraints on a finite number of known 
rotationally invariant variables constructed from a Gaussian (or weakly non-Gaussian) random field. 
We will demonstrate, in a systematic way, that the Lagrangian bias parameters are the coefficients of the 
expansion of the constraints in the basis of orthonormal polynomials associated with the different variables.
These polynomials need not be multivariate Hermite, as rotational invariance may require to construct 
nonlinear functionals of the (nearly) Gaussian field.
Even so, since the coefficients are ensemble averages of the same polynomials, no renormalization is needed. 

Furthermore, these bias factors are still peak-background split biases, in the sense that they
encode the response of the tracer number density to long-wavelength perturbations.
We will also show that some of these variables behave like dynamical angles (in contrast to geometrical 
angles), and take advantage of this to re-derive the well-known BBKS formula for the average peak number 
density (see \cite{bardeen/bond/etal:1986}) in a much simpler way. 
However, the gravitational evolution or the connection to the integrated perturbation theory (iPT)
\cite{matsubara:2011} will be presented elsewhere.

As an application of our formalism, we compute the amplitude of the non-Gaussian bias imprinted by a 
local primordial trispectrum. This effect has already been explored in 
\cite{desjacques/seljak:2010,desjacques/jeong/schmidt:2011,smith/ferraro/loverde:2012,yokoyama/matsubara:2013}. 
Here, we also point out that, for weak primordial non-Gaussianity, the third-order Lagrangian bias 
parameters must satisfy a consistency relation similar to that obtained in 
\cite{desjacques/gong/riotto:2013,biagetti/desjacques:2015}.
This consistency relation cannot be fulfilled by models of halo bias in which the Lagrangian clustering 
depends on the local mass density solely.

The paper is organized as follows. We begin with a short review the peak formalism and the approach of D13 
in \S\ref{sec:peaktheory}. In \S\ref{sec:determinant}, we clarify the dependence of peak clustering 
on angular dynamical degrees of freedom such as the determinant, and present a simple derivation of the 
BBKS formula for the average peak number density. In \S\ref{sec:multivariatebias}, we describe our 
general methodology for constructing Lagrangian perturbative bias expansions. In \S\ref{sec:PNG}, we 
apply our method to the non-Gaussian bias induced by a primordial trispectrum. 
We conclude in \S\ref{sec:conclusion}.

\section{A brief review of peak theory}
\label{sec:peaktheory}

The peak formalism was pioneered by \citep{bardeen/bond/etal:1986}, and has been 
extended in a number of recent studies 
\citep{desjacques:2008,desjacques/crocce/etal:2010,desjacques/sheth:2010,desjacques:2013}.
To ensure that the initial density peaks are in one-to-one correspondence with isolated (parent)
halos, an additional constraint is imposed on the slope of the filtered density field, and results
in the so-called excursion set peaks (hereafter ESP) 
\cite{paranjape/sheth:2012,paranjape/sheth/desjacques:2013}.
For simplicity, we will stick to the original model of \cite{bardeen/bond/etal:1986}, i.e. the BBKS peaks. 
However, our approach can be generalized to include more realistic constraints, such as excursion set peaks, 
a dependence on the initial tidal shear \cite{bond/myers:1996,sheth/chan/scoccimarro:2013}, etc.
Furthermore, we will assume Gaussian initial conditions throughout this Section, so that probability density 
functions (PDF) are given by multivariate Normal $\mathcal{N}$.

\subsection{Variables}

To enforce the peak constraint, knowledge of the linear density field smoothed on the halo mass scale 
$R\propto M^{1/3}$  and of its derivatives is required. 
For subsequent use, we introduce the following quantities built from the convolution 
$\delta_R(\mb{x})\equiv (W_R\star \delta_L)(\vx)$ of the linear matter density field $\delta_L$ with a with a 
filter $W_R$:
\begin{gather}
\label{eq:variables}
\nu(\vx) = \frac{1}{\sigma_0}\delta_R(\vx)\; , \qquad
\boldsymbol{\eta}(\vx) = \frac{1}{\sigma_1}\grad\delta_R(\vx)\;, \\
\zeta_{ij} (\vx) = \frac{1}{\sigma_{2}}\partial_i \partial_j \delta_R(\vx) \;. 
\end{gather}
These variables are normalized to unity by the spectral moments $\sigma_j$ of the linear mass density 
field
\begin{equation}
\sigma_j^2 \equiv \frac{1}{2\pi^2} \int_0^\infty \!\!dk\, k^{2(j+1)}\tilde{W}_R^2(k)\,P_L(k) \;,
\end{equation}
where $P_L(k)$ is the linear power spectrum and $\tilde{W}_R(k)$ is the Fourier transform of the filter $W_R$, which should decay sufficiently 
rapidly so that the integrals converge. 

Clearly, $\nu(\mathbf{x})$ is invariant under spatial rotations, and so is the square gradient of 
the density field 
\begin{equation}
  \label{eq:eta}
  \eta^2(\mathbf{x}) 
  = \frac{1}{\sigma^2_{1}} \big(\grad\delta_R\big)^2\!(\vx) \;.
\end{equation}
Three more rotational invariants can be constructed from the Hessian matrix $\zeta_{ij}$. 
In the notation of \cite{gay/pichon/pogosyan:2012}, we have
\begin{equation}
  J_1= -\tr(\zeta_{ij})\;, \quad
  J_2= \frac{3}{2} \tr(\bar\zeta_{ij}^2)\;, \quad 
  J_3=\frac{9}{2}\tr (\bar\zeta_{ij}^3) \;,
\end{equation}
where $\bar\zeta_{ij}\equiv \zeta_{ij}+ \delta_{ij}J_1/3$ are the component of the traceless part of the Hessian
~\footnote{In the notation of D13, these variables are $u\equiv J_1$ and $\zeta^2=J_2$.}.
These invariants are related to the coefficients of the characteristic polynomial det$(\zeta-\lambda\vii)$, where 
$\zeta$ is the matrix of second derivatives and $\vii$ is the $3\times 3$ identity matrix.
In particular, the determinant of the Hessian matrix is related to the rotational invariants by
\begin{equation}
  J_3(\vx) = \frac{1}{2} J_1^3(\vx) - \frac{3}{2} J_1(\vx) J_2(\vx) 
  + \frac{27}{2}\det(\zeta_{ij})(\vx) \;,
\label{eq:J3}
\end{equation}
owing to Newton's identities.

The Fourier space expression of these invariants is
\begin{align}
\nu(\vx) &= \frac{1}{\sigma_0} 
\int\!\!\frac{d^3 k}{(2\pi)^3}\, \delta_R(\vk) e^{i\vk\cdot\vx} \;,\label{eq:nuFT} \\
 J_1(\vx) &= \frac{1}{\sigma_2} 
\int\!\!\frac{d^3 k}{(2\pi)^3}\, \delta_R(\vk) k^2 e^{i\vk\cdot\vx} \;, \label{eq:uFT}
\end{align}
for the peak height and curvature, 
\begin{align}
\eta^2(\vx) &= -\frac{1}{\sigma^2_1} 
\iint\!\frac{d^3 k_1}{(2\pi)^3}\frac{d^3k_2}{(2\pi)^3}\, \delta_R(\vk_1) \delta_R(\vk_2) \nonumber \\
& \qquad \times \left(\vk_1\cdot\vk_2\right) \,  e^{i(\vk_1+\vk_2)\cdot\vx} \;,
\label{eq:etaFT}
\end{align}
for the square gradient, and
\begin{align}
  J_2(\mb{x}) &= \frac{3}{2 \sigma^2_2} \iint\frac{d^3 k_1}{(2\pi)^3} \frac{d^3k_2}{(2\pi)^3}\, 
\delta_R(\mb{k}_1) \delta_R(\mb{k}_2) \notag \\
 & \qquad \times
\left[\left(\mb{k}_1\cdot\mb{k}_2\right)^2  - \frac{1}{3}k_1^2 k_2^2\right]   
e^{i(\mb{k}_1+\mb{k}_2)\cdot\mb{x}} \;,
\end{align}
\begin{widetext}
\begin{align}
J_3(\mb{x}) &= -\frac{9}{2 \sigma^3_2} \iiint\frac{d^3 k_1}{(2\pi)^3} \frac{d^3k_2}{(2\pi)^3}
\frac{d^3k_3}{(2\pi)^3}\, 
\delta_R(\mb{k}_1) \delta_R(\mb{k}_2) \delta_R(\mb{k}_3)
\bigg\{\left(\mb{k}_1\cdot\mb{k}_2\right)\left(\mb{k}_2\cdot\mb{k}_3\right)
\left(\mb{k}_3\cdot\mb{k}_1\right)  \\ 
&\qquad 
-\frac{1}{3}\Big[\left(\mb{k}_1\cdot\mb{k}_2\right)^2 k_3^2 + \mbox{2 cyc.}\Big]
+\frac{2}{9}k_1^2 k_2^2 k_3^2\bigg\} e^{i(\mb{k}_1+\mb{k}_2+\mb{k}_3)\cdot\mb{x}} \;, \nonumber
\end{align}
\end{widetext}
for the second- and third-order invariants constructed out of $\bar\zeta_{ij}$. 

We will collectively refer to the Gaussian variables as 
\begin{equation}
  \vy \equiv 
  \big\{\nu(\vx), J_1(\vx), \sqrt{3}\eta_i(\vx),
  \sqrt{5}\bar\zeta_{ij}(\vx) \big\} \;,
\end{equation}
and call $\mathcal{N}(\vy)$ their $N$-variate normal PDF. 
Finally, we will group all  the (not necessarily Gaussian) rotational invariants as
\begin{equation}
  \vw \equiv
  \big\{\nu(\vx), J_1(\vx), 3\eta^2(\vx), 5J_2(\vx),J_3(\vx) \big\} \;,
\end{equation}
and call $P(\vw)$ their PDF.
In the case of BBKS peaks, $\vy$ is a 10-dimensional list  containing $\nu$, $J_1$, the three components of 
$\boldsymbol{\eta}$ and the five independent entries of $\bar\zeta_{ij}$. To describe ESP, $\vy$ should 
also include the slope $\mu\sim -d\delta_R/dR$ of the excursion set trajectories. 

\subsection{Perturbative peak bias expansion}
\label{sec:ELBE}

As explained in BBKS, the peak point process in three dimensions has the {\it localized} number density  
\begin{equation}
\cpk(\vx) = \frac{3^{3/2}}{R^3_\star}
\bigl\lvert\mbox{det}\zeta_{ij}\bigr\lvert\,
\delta_D(\boldsymbol{\eta})\Theta_H(\lambda_3)\,
\delta_D(\nu - \nu_c) \;.
\label{eq:npk}
\end{equation}
Here, $\nu_c = \delta_c / \sigma_0$, where $\delta_c$ is the linear density contrast for spherical collapse, $R_{\star} = \sqrt{3} (\sigma_1 / \sigma_2)$ is the characteristic radius of a peak, $\lambda_3$ is the smallest eigenvalue of $-\zeta_{ij}$, and $\Theta_H$ is the Heaviside step function. 
We will refer to Eq.(\ref{eq:npk}) as the peak constraint. 
It returns the number of peaks of height $\nu_c$ (either 0 or 1) in the infinitesimal volume $d^3x$, divided by $d^3x$. The point process $\cpk(\vx)$ depends on $\vx$ only through $\vy$, a set of continuous Gaussian fields evaluated at the same position $\vx$, and when evaluating statistics we will often denote it $\cpk(\vy)$.

The average peak number density is 
\begin{equation}
\label{eq:bnpk}
\bnpk = \int \!\! d\vy\,\cpk(\vy)\,{\cal N}(\vy) \;,
\end{equation}
where ${\cal N}(\vy)$ is the 10-variate normal PDF of $\vy$, and is also computed in BBKS. We will come back to this 
shortly. 

The peak overdensity is defined locally as
\begin{equation}
\label{eq:locpeak}
  \dpk (\vx) \equiv \frac{\cpk(\vx)}{\bnpk} - 1 \;.
\end{equation}
When computing correlation functions, D13 has argued that $\dpk (\mathbf{x})$ can be replaced by its 
{\it effective} or {\it mean-field} perturbative peak bias expansion (we follow the notation advocated in this paper, 
rather than D13)
\begin{align}
\dpk (\mathbf{x})  \simeq &\,\sigma_{0} b_{10} \nu(\mathbf{x}) 
+ \sigma_{2} b_{01} J_1(\mathbf{x}) + \frac{1}{2} \sigma^{2}_{0} b_{20} \nu^2(\mathbf{x}) 
\label{eq:biasExp} \\
& + \sigma_{0} \sigma_{2} b_{11} \nu(\mathbf{x}) J_1(\mathbf{x}) 
+ \frac{1}{2} \sigma^{2}_{2} b_{02} J_1^2(\mathbf{x}) \nonumber \\
&  + \bigg. \sigma^{2}_{1} \chi_1 \eta^2(\mathbf{x}) +\sigma^{2}_{2} \omega_{10} 
J_2(\mathbf{x}) + \dots  \nonumber
\end{align}
which bears similarities to the standard local bias expansion (see e.g. \cite{fry/gaztanaga:1993}). 
This effective overabundance satisfies $\dpk(\vx)\geq -1$ only if all the terms in the infinite series expansion are 
included.
When expanded in this form,  $\dpk(\vx)$ is generally {\it not} a count-in-cell (measurable) density, but rather a 
mean-field (or most probable) peak overdensity given a certain realisation of the density field 
and its derivatives. 
In fact, Eq.(\ref{eq:locpeak}) is equivalent to Eq.(\ref{eq:biasExp}) only if the realisation has infinite volume, in 
which case ergodicity ensures that the series coefficients or bias factors are the ensemble averages
\begin{align}
\sigma^i_0 \sigma^j_2 b_{ij} &= 
\frac{1}{\bnpk} \int\!\! d\vy\,\cpk(\vy)\, H_{ij}(\nu,J_1) {\cal N}(\vy) 
\label{eq:biasfactors} \\
\sigma_1^{2q}\chi_q &= \frac{(-1)^q}{\bnpk}
\int\!\! d\vy\,\cpk(\vy)\,L^{(1/2)}_q\!\left(\frac{3 \eta^2}{2}\right)  {\cal N}(\vy) 
\nonumber \\
\sigma_2^{2k}\omega_{k0} &= \frac{(-1)^k}{\bnpk}
\int\!\! d\vy\,\cpk(\vy)\,L^{(3/2)}_k\!\left(\frac{5 J_2}{2}\right)  {\cal N}(\vy)
\nonumber \;.
\end{align}
The multiplicative factors of $\sigma_j$ arise from the bias factors being defined relative to the 
unnormalized fields $\delta(\vx)$, $(\grad\delta)^2(\vx)$ etc.

Finally, we have also introduced the bivariate Hermite polynomials
\begin{equation}
  H_{ij}(\nu,J_1) = \frac{1}{\mathcal{N}(\nu,J_1)} \!
  \left(-\frac{\partial}{\partial\nu}\right)^{\!i}\!\! 
  \left(-\frac{\partial}{\partial J_1}\right)^{\!j} \!
  \mathcal{N}(\nu,J_1) \;,
\label{eq:H}
\end{equation}
and the generalized Laguerre polynomials
\begin{equation}
  L^{(\alpha)}_n (x) =
  \frac{x^{-\alpha}e^x}{n!} \frac{d^n}{dx^n}
  \left(e^{-x}x^{n+\alpha}\right) \;.
\label{eq:L}
\end{equation}
Note also that, in principle, it is possible to weight peaks such as to cancel the complicated weight brought 
by the peak constraint \cite{verde/jimenez/etal:2014}, but it is not obvious how to accomplish this with real 
data.

Let us make two comments. Firstly, D13 argued that the determinant, hence $J_3$, does not appear in the 
perturbative peak bias expansion Eq.~(\ref{eq:biasExp}) because det$(\zeta_{ij})$ does not appear in the 
multivariate normal ${\cal N}(\vy)$.
His statement, however, was backed up with a perturbative calculation of the peak-peak correlation 
function $\xpk(r)$ at second order only. We will revisit this question and show that det$(\zeta_{ij})$ 
is present in ${\cal N}(\vy)$. Therefore, det$(\zeta_{ij})$ appears in the expansion of $\dpk(\mathbf{x})$ with 
its own bias coefficients.

Secondly, D13 speculated that the powers appearing in Eq.\eqref{eq:biasExp}  should actually be replaced by 
orthogonal polynomials, e.g. $J_2(\vx)$ should be replaced by $L_1^{3/2}(5J_2(\vx)/2)$, quadratic terms 
in $\nu(\vx)$ and $J_1(\vx)$ by Hermite polynomials etc. This would ensure that all the terms involving zero-lag 
correlators -- like $\avg{\nu^2(\vx)}$ or $\avg{J_2(\vx)}$ -- automatically cancel out in the calculation 
of $\xpk(r)$, which becomes considerably simpler. Otherwise, this cancellation must be enforced explicitly,
as was done in D13 or in \cite{schmidt/jeong/desjacques:2013} for instance. 
We will hereafter clarify this point, and illustrate how the cancellation of zero-lag terms becomes the 
guiding principle in the choice of the appropriate orthogonal system.

\section{Angles and determinant}
\label{sec:determinant}

In this Section, we clarify the dependence of peak clustering statistics on the determinantal variable $J_3$, 
and present an easy derivation of the well-known BBKS formula for the average peak number density.

\subsection{Joint probability density}

For the purpose of diagonalizing the PDF $\mathcal{N}(\vy)$, we introduce the variable
\begin{equation}
\label{eq:z}
z\equiv \frac{J_1-\gamma_1 \nu}{\sqrt{1-\gamma_1^2}} \;,
\end{equation} 
which does not correlate with $\nu$. 
Here, $\gamma_1 \equiv \sigma^2_1/\sigma_0 \sigma_2$ is the correlation coefficient of the bivariate Normal
${\cal N}(\nu,J_1)$.
The probability density ${\cal N}(\vy)$ thus becomes
\begin{align}
\label{eq:PG}
P(&\nu,z,3\eta^2,5J_2,J_3,\vomg) \\ 
& = \frac{5\sqrt{5}}{8\pi^2\sqrt{3}} \sqrt{\eta^2}
\exp\left(-\frac{1}{2}\nu^2-\frac{1}{2}z^2-\frac{3}{2}\eta^2-\frac{5}{2}J_2\right) \nonumber \\
&\qquad \times P(\vomg) \nonumber \;,
\end{align}
where $\vomg$ is a vector of five angular variables, two of which characterize the direction of the 
gradient $\grad\delta_R$, and the remaining three describe the orientation of the principal axis frame of 
$\zeta_{ij}$ (Euler angles).  $P(\vomg)$ is the corresponding PDF.
The reason for using $3\eta^2$ and $5J_2$ will become apparent shortly. 
Importantly, we always have $J_2>0$ and $J_3^2\leq (J_2)^3$ for any symmetric matrix. 
Therefore, $\mathcal{N}(\vy)$ is normalized such that
\begin{multline}
1 = \int d\vomg
\int_{-\infty}^{+\infty}\!\!d\nu\int_{-\infty}^{+\infty}\!\!dz\int_0^\infty\!\!d(3\eta^2) 
\int_0^\infty\!\!d(5J_2)\\
\times \int_{-J_2^{3/2}}^{+J_2^{3/2}}\!\!dJ_3\, P(\nu,z,3\eta^2,5J_2,J_3,\vomg) \;,
\end{multline}
even though it does not explicitly depend on $J_3$. Clearly, the integral over $J_3$ is trivial and 
results in $5J_2$ being $\chi^2$-distributed with 5 degrees of freedom. These 5 d.o.f. correspond 
to the 5 independent components of $\bar\zeta_{ij}$. This is the reason for writing $5J_2$. 
Similarly, one can easily show that $3\eta^2$ is $\chi^2$-distributed with 3 d.o.f.

To emphasize the point that $J_3$ is an ``angular'' variable, we introduce $x_3\equiv J_3/(J_2)^{3/2}$.
As a result, $\mathcal{N}(\vy)$ transforms into
\begin{align}
  \label{eq:newfactorization}
  \mathcal{N}(\vy) & d\vy
  = {\cal N}(\nu) d\nu
  \,{\cal N}(z) dz
  \,\chi^2_3(3\eta^2) d(3\eta^2) \\
  & \times \chi^2_5(5J_2) d(5J_2)
  \frac{1}{2}\Theta_H\!(1-x_3^2)dx_3
  P(\vomg) d\vomg \nonumber \;.
\end{align}
In the above expression, $\chi^2_k (x)$ is a $\chi^2$ distribution with $k$ degrees of freedom,
\begin{equation}
\chi^2_k (x) = \frac{1}{2^{k/2} \Gamma(k/2)} x^{k/2 -1} e^{-x/2} \;.
\label{eq:chi}
\end{equation}
Clearly, $x_3$ has uniform probability density in the range $-1\leq x_3\leq +1$ and, thus, behaves as 
(the cosine of) an angle. However, this angle has nothing to do with spatial rotations, but with the 
solution of the cubic equation for the eigenvalues of $\zeta_{ij}$: it is therefore itself a rotational 
invariant.
We will in fact see that, unlike $\vomg$ which do not contribute to the clustering, $J_3(\vx)$ actually 
does and, therefore, $x_3$ does not describe a rotation in space.

Finally, the dependence of the upper and lower bound of the $J_3$-integral on the value of $J_2$ implies 
that both variables are correlated. Consequently, one should expect new, double-index valued bias 
coefficients for both $5J_2$ and $J_3$ such that they reduce to $\omega_{k0}$ when the $J_3$-index is 
zero.

\subsection{Determinant and peak constraint: a simple derivation of the BBKS formula}

The peak constraint implies that all three eigenvalues of the Hessian $\zeta_{ij}$ be negative. 
In terms of the rotational invariant, the restriction to local maxima of the density field translates 
into the conditions $J_1>0$, $J_2<J_1^2$ and $J_3<(J_1/2)(J_1^2-3J_2)$. 
Taking into account the symmetry of $\zeta_{ij}$, the last condition implies that $x_3$ must satisfy
\begin{equation}
  -1<x_3<\mbox{min}\big[1,(y/2)(y^2-3)\big] \;,
\label{eq:detconstraint}
\end{equation}
where $y\equiv J_1/\sqrt{J_2}$. This splits the parameter space into two different regions depending on 
whether the inequality $(y/2)(y^2-3)<1$ holds. For $0<J_2<J_1^2/4$, one finds $-1<x_3<+1$ whereas, for 
$J_1^2/4<J_2<J_1^2$, the more stringent constraint $-1<x_3<(y/2)(y^2-3)$ applies. 
Therefore, the multiplicative factor of $\Theta_H(\lambda_3)$ in the localized peak number density 
Eq.(\ref{eq:npk}) also reads
\begin{align}
\Theta_H(\lambda_3)& = \Theta_H(J_1)
\bigg\{\Theta_H\big(J_1^2/4-J_2\big) + \Theta_H\big(J_2-J_1^2/4\big)
\nonumber \\
& \times \Theta_H\big(J_1^2-J_2\big)\, \Theta_H\big(y^3/2-3y/2-x_3\big)\bigg\} \;.
\end{align}
We are now in the position to derive the well-known BBKS formula for the average peak number density $\bnpk$
in a very simple way. 

To compute $\bnpk$, one usually expresses the measure $d(5J_2)$ in terms of the ellipticity $v$, the 
prolateness $w$ and three Euler angles so that, in these new variables, $J_2=3v^2+w^2$. 
The calculation then proceeds along the lines of BBKS.
However, this change of variable is, in fact, unnecessary as the calculation can be explicitly carried out 
in the variables $J_1$, $5J_2$ and $x_3$ on imposing the aforementioned conditions.

To illustrate this point, we begin by rewriting the determinant $|\det(\zeta_{ij})|$ in Eq.\ref{eq:npk}
using Eq.\ref{eq:J3}. Introducing $s\equiv 5J_2$, this can be written as
\begin{align}
|\det(\zeta_{ij})|
&= \frac{1}{27}\left(J_1^3-\frac{3}{5}s J_1 - \frac{2}{5^{3/2}}s^{3/2}x_3\right) \;,
\end{align}
since $\det(\zeta_{ij})$ is always negative for density maxima. The integral over the variables $5J_2$ and 
$x_3$ becomes 
\begin{widetext}
\begin{align}
\int\!\!ds\, \chi_5^2(s) & \int\!\!dx_3\,\frac{1}{2}\Theta_H(1-x_3^2)\, 
\big\lvert\mbox{det}(\zeta_{ij})\big\lvert\,\Theta_H(\lambda_3) \\
&= \frac{1}{2^{7/2}3^3\Gamma(5/2)}\biggl\{\int_0^{5J_1^2/4}\!\!ds\int_{-1}^{+1}\!\!dx_3+
\int_{5J_1^2/4}^{5J_1^2}\!\!ds\int_{-1}^{\sqrt{5/s}(J_1/2)(5J_1^2/s-3)}\!\!dx_3\biggr\} \nonumber \\
&\qquad \times \left(J_1^3-\frac{3}{5}s J_1 - \frac{2}{5^{3/2}}s^{3/2}x_3\right)  s^{3/2} e^{-s/2} \;, \nonumber
\end{align}
and can be computed straightforwardly with the aid of the (lower) incomplete Gamma function 
$\gamma(\lambda,s)=\int_0^s\!ds'\, (s')^{\lambda-1} e^{-s'}$. Taking into account two additional multiplicative
factors of $3^{3/2}$, one arising from $\cpk(\vy)$ and the other from the integral over 
$\chi_3^2(3\eta^2)\delta_D(\veta)$, we find
\begin{align}
3^3\int\!\!ds\, \chi_5^2(s) & \int\!\!dx_3\,\frac{1}{2}\Theta_H(1-x_3^2)\, 
\big\lvert\mbox{det}(\zeta_{ij})\big\lvert\,\Theta_H(\lambda_3) \\
&= \sqrt{\frac{2}{5\pi}}\left[\left(\frac{J_1^2}{2}-\frac{8}{5}\right)e^{-5J_1^2/2}
+\left(\frac{31}{4}J_1^2+\frac{8}{5}\right) e^{-5J_1^2/8}\right]+\frac{1}{2}\left(J_1^3-3J_1\right) \nonumber \\
&\qquad \times 
\left[\mbox{Erf}\!\left(\sqrt{\frac{5}{2}}J_1\right)+\mbox{Erf}\!\left(\sqrt{\frac{5}{2}}\frac{J_1}{2}\right)\right]
\equiv f(J_1) \nonumber \;,
\end{align}
\end{widetext}
which is precisely the function $f(J_1)$ defined in BBKS. 
The rest of the calculation is trivial, and we immediately recover their well-known expression for the average 
peak abundance $\bnpk(\nu_c)$,
\begin{equation}
  \bnpk(\nu_c) = \frac{1}{(2\pi)^2 R_\star^3} 
  G_0(\gamma_1,\gamma_1\nu_c) e^{-\nu_c^2/2}\;,
\end{equation} 
where the function $G_0$ is
\begin{align}
G_0(\gamma_1,\omega)=\int_0^\infty\!\!du\,f(u)\,
\frac{e^{-(u-\omega)^2/2(1-\gamma_1^2)}}{\sqrt{2\pi(1-\gamma_1^2)}}\;.
\end{align}
Note, however, that our derivation is far simpler than that presented in BBKS.

\section{Constructing perturbative bias expansions}
\label{sec:multivariatebias}

We will now generalize the calculation of D13. 
Since the initial conditions are nearly Gaussian, it is sensible to treat primordial non-Gaussianity (PNG) as 
a small perturbation. 
The key point is that, while the effective density of biased tracers $n_\text{tr}(\vx)$ is a function of 
(nearly) Gaussian variables that transform like scalars, vectors, tensors etc. under a global rotation, it is 
itself invariant under this transformation. 
Therefore, $n_\text{tr}(\vx)$ can generically be written as a series expansion in orthogonal polynomials,  
which are invariant under global rotations.
These polynomials remove all the zero-lag correlators, like Hermite polynomials do for scalar Gaussian variables. 
This ensures that the polynomials remain orthogonal also at finite separation, and considerably simplifies the 
calculation of correlation functions.

Some models, like the BBKS peaks considered here, have an additional $SO(3)$ symmetry due to the fact that the 
constraint defining the tracer is independent of the angle of (one of) the vector(s) with the eigenvectors of 
(one of) the tensor(s).
In this particular case, the orthogonal polynomials will be invariant under $SO(3)\times SO(3)$. 
We emphasize, however, that this symmetry is completely independent of  the correlation structure of the Gaussian 
variables. 
If, for instance, the constraint involved the alignment angle $\eta_i\bar\zeta_{ij}\eta_j$, this would break the second 
$SO(3)$ symmetry even if $\bm{\eta}$ and $\bar\zeta$ remain statistically uncorrelated.

Our generalization is based exclusively on the symmetries of the constraint defining the biased tracers. Hence, 
it can be readily applied to models more sophisticated than the BBKS or ESP peaks. 

\subsection{General methodology}

The methodology is straightforward. First, we construct a list $\vy(\vx)$ of all the Gaussian variables which 
$n_\text{tr}$ depends on, ordered according to their rank (scalar, vector, tensor...). These are labeled by a 
greek index $\alpha$, and normalized so that $\avg{y_\alpha^2}=1$.
The ``localized number density'' of tracers is a point process specified by a ``selection function'' 
$n_\text{tr}(\vy)$ that sets a number of constraints on $\vy$. For instance, we have $n_\text{tr}(\vy)=\cpk(\vy)$ 
in the case of the BBKS peaks discussed in Sec. \S\ref{sec:determinant}.
Because of rotational invariance, $n_\text{tr}(\vy)$ depends on $\vy$ only through a smaller set of invariant 
quantities  $\vw$. 
The probability distribution factorizes into $\mathcal{N}(\vy)d\vy = P(\vw) d\vw \times  P(\vomg) d\vomg$, 
where $\vomg$ is a set of geometric angles, while $n_\text{tr}(\vy)=n_\text{tr}(\vw)$.

The effective abundance $n_\text{tr}(\vx)$ of biased tracers is conveniently expanded in the multivariate 
Hermite polynomials \cite{szalay:1988}, defined as
\begin{align}
  \label{eq:multihermite}
  H^{(n)}_{\alpha_1,...,\alpha_n}(\vy) &\equiv
  \frac{(-1)^n}{\mathcal{N}(\vy)}
  \frac{\pd^n\mathcal{N}(\vy)}{\pd y_{\alpha_1} \dots \pd y_{\alpha_n}}\;.
\end{align}
For a generic covariance matrix $\avg{y_\alpha y_\beta}$ these polynomials are orthogonal, relative to the weight 
$\mathcal{N}(\vy)$,  to their duals
\begin{align}
  H^{\star(n)}_{\alpha_1,...,\alpha_n}(\vy)
  &\equiv \sum_{\beta_1,\dots,\beta_n}
  \big\langle y_{\alpha_1}y_{\beta_1}\big\rangle\cdots\big\langle 
  y_{\alpha_n}y_{\beta_n}\big\rangle \\
  & \qquad \times  H^{(n)}_{\beta_1,...,\beta_n}(\vy)\;. \nonumber
\end{align}
This property allows to expand $n_\text{tr}(\vx) = n_\text{tr}\big(\vy(\vx)\big)$ as
\begin{equation}
  \label{eq:hermbias}
  n_\text{tr}(\vx) =   \sum_{n=0}^\infty
  \frac{1}{n!}
  \big\langle n_\text{tr} H^{(n)}_{\alpha_1,...,\alpha_n}\big\rangle
  H^{\star(n)}_{\alpha_1,...,\alpha_n}\big(\vy(\vx)\big)\,,
\end{equation}
where the brackets denote the ensemble average {\it at a single location} over all the possible realisations of $\vy$, 
weighted by $\mathcal{N}(\vy)$ and subject to the set of constraints $n_\text{tr}(\vy)$ that define the biased process. 

Although formally correct (see \cite{szalay:1988} for discussion), this expansion flattens out the $SO(3)$ structure 
of the coefficients, hiding the symmetries that arise because of rotational invariance. 
Since some $\pd/\pd y_\alpha$ stand for $\pd/\pd\eta_i$, $\pd/\pd\bar\zeta_{ij}$ etc., the Hermite polynomials 
\eqref{eq:multihermite} also carry vector, tensor etc. indices that transform under rotation accordingly. 
When the constraint $\mathcal{C(\vy)}$ satisfies a more restrictive $SO(3) \times SO(3)$ symmetry, these indices are 
contracted by Kronecker symbols in the coefficients of \eqref{eq:hermbias}.
Therefore, what will actually appear in the expansion are the polynomials \eqref{eq:multihermite} averaged over 
different directions.
These averages may be computed taking all possible ways of contracting the indices, with their combinatorial weight. 
Equivalently, but more conveniently, one may orthonormalize with the Gram-Schmidt procedure the polynomials obtained 
using invariant differential operators in \eqref{eq:multihermite}. 

The resulting polynomials are of order $n$ in the Gaussian field, but are invariant by construction, and therefore 
must depend only on $\vw$. 
For convenience, we normalize them so as to include the $1/n!$ factor in each term of the series expansion 
\eqref{eq:hermbias} into their definition. 
We call these normalized polynomials $\tilde{O}_{\vn}(\vw)$, where $\vn$ is a list of indices. 
Each index $n_i$ corresponds to an invariant $w_i\in \vw$, and comes with a weight $\varpi_i$ such that 
$\sum_i n_i \varpi_i=n$. This weight corresponds to the degree (i.e. powers of the density field) of $w_i$.
The normalized polynomials are orthonormal relative to the weight $P(\vw)$, and thus satisfy 
\begin{equation}
  \int d\vw \, \tilde O_{\vn}(\vw) \tilde O_{\vm}^\star(\vw) P(\vw)
  = \delta_{\vn,\vm} \,,
\end{equation}
where $\vn$ and $\vm$ are lists of integers, and $\delta_{\vn,\vm}$ is a multi-dimensional Kronecker symbol.

The effective or mean-field overdensity $\delta_\mathrm{tr}(\vx)$ of the biased tracers is now given by the expansion
\begin{align}
\label{eq:generalizedbias}
\delta_\mathrm{tr}(\vx) &= \sum_{\vn\neq{\bf 0}}
\tilde c_\vn\, \tilde O^\star_{\vn}(\vw) \;.
\end{align}
The normalized bias coefficients are
\begin{equation}
  \tilde c_\vn \equiv \frac{1}{\bar n_\mathrm{tr}}\big\langle n_\text{tr}(\vw)\,\tilde O_{\vn}(\vw)\big\rangle \;, 
\end{equation}
where $\bar n_\mathrm{tr} = \avg{n_\text{tr}}$ is the average number density of tracers. 
We can reexpress the perturbative bias expansion on collecting all the terms of order $n$ in the fields. In Fourier 
space, this takes the form
\begin{align}
\delta_\text{tr}(\vk) &= \sum_{n=1}^\infty \frac{1}{n!}
\int\!\!\frac{d^3k_1}{(2\pi)^3}\dots\frac{d^3k_n}{(2\pi)^n}\,
c_n^L(\vk_1,\dots,\vk_n) \\
& \qquad \times \delta(\vk_1)\dots\delta(\vk_n) \delta_D(\vk_{1...n}-\vk)
\nonumber \;,
\end{align}
where $\vk_{1...n}=\vk_1+\dots+\vk_n$ and $c_n^L(\vk_1,\dots,\vk_n)$ is the $n$th-order Fourier space Lagrangian
bias. We use the notation of \cite{matsubara:2011} to emphasize that the $c_n^L$ precisely are the renormalized
bias functions of iPT. The connection between peak theory and the iPT will be discussed elsewhere.

Finally, like the multivariate Hermite polynomials they are constructed from, the polynomials 
$\tilde O^\star_{\vn}(\vw)$ also have the property of removing all zero-lag Gaussian correlators from correlation 
functions: they will thus remain orthogonal also at finite separation. 
Their correlation function explicitly reads
\begin{equation}
  \Big\langle\tilde O^\star_{\vn}\big(\vw(\vx_1)\big)
  \tilde O^\star_{\vm}\big(\vw(\vx_2)\big)\Big\rangle
  = \delta_{\vn,\vm} C_\vn(\vx_1,\vx_2)\;,
\end{equation}
where $C_\vn(\vx_1,\vx_2)$ contains only rotationally invariant combinations of the two-point correlation functions 
of $\vy$ {\it at different locations}. Consequently,
\begin{equation}
  \Big\langle\delta_\mathrm{tr}(\vx_1)\delta_\mathrm{tr}(\vx_2)\Big\rangle =
  \sum_{\vn} \tilde c_\vn^2 C_\vn(\vx_1,\vx_2) \;.
\end{equation}
Similarly, the connected $N$-point correlation functions of the biased tracers can now be computed perturbatively 
upon evaluating the ensemble averages
\begin{equation}
\Bigl\langle\delta_\text{tr}(\vx_1)\times \dots\times \delta_\text{tr}(\vx_N)\Big\rangle \;.
\end{equation}
We will now apply this symmetry-based approach to BBKS peaks and derive the corresponding form of $\tilde O_\vn(\vw)$ 
at all orders. We will also show how to compute their correlation functions at finite separation.

\subsection{Application to BBKS peaks}

The constraint for BBKS peaks involves up to second derivatives of the density field, such that the relevant 10 Gaussian
variables are $\vy=\{\nu(\vx),J_1(\vx),\sqrt{3}\eta_i(\vx),\sqrt{5}\bar\zeta_{ij}(\vx)\}$. As a result, $\vomg$ contains
5 angles, and $\vw$ splits into three subsets of 1-point correlated variables: $(\nu,J_1)$, $(3\eta^2)$ and $(5J_2,J_3)$.
As stated above, our procedure is completely general and can be easily extended to more realistic models. In the case
of ESP peaks for instance, we would simply add the normalized slope $\mu\equiv -(d\delta/dR)/\sqrt{\avg{(d\delta/dR)^2}}$ 
of the excursion set trajectories ($R$ is the smoothing radius) to the set of variables $\vy$.

\subsubsection{Polynomials}

The differential operators that preserve the $SO(3) \times SO(3)$ invariance are 
$\pd_\nu \equiv \pd/\pd\nu$, 
$\pd_{J_1} \equiv \pd/\pd J_1$, and
\begin{align}
  \nabla^2_\eta &\equiv
  \frac{1}{3}\grad_\eta\cdot\grad_\eta
  = \frac{1}{3}\frac{\pd}{\pd\eta_i}\frac{\pd}{\pd\eta_i}\,,
  \\
  D_2 & \equiv \frac{1}{5}\tr[(\pd/\pd\bar\zeta)^2]
  = \frac{1}{5}\frac{\pd}{\pd\bar\zeta_{ij}}
  \frac{\pd}{\pd\bar\zeta_{ji}}\,,
  \\
  D_3 & \equiv \frac{1}{5^{3/2}}\tr[(\pd/\pd\bar\zeta)^3]
  = \frac{1}{5^{3/2}} \frac{\pd}{\pd\bar\zeta_{ij}}
  \frac{\pd}{\pd\bar\zeta_{jk}} \frac{\pd}{\pd\bar\zeta_{ki}}\,.
\end{align}
Higher order operators like $D_4$, $D_5$, and so on, should not be included, because a 3$\times$3 traceless matrix 
(even of derivatives) has only two rotational invariants. 
Mixed operators such as $(\pd/\pd\eta_i)(\pd/\pd\bar\zeta_{ij})(\pd/\pd\eta_j)$ are not allowed, because not invariant 
when $\eta_i$ and $\bar\zeta_{ij}$ transform under different rotations.

For the purpose of this derivation, we need to introduce the Legendre polynomials
\begin{equation}
  P_m(x) = \frac{1}{2^mm!}\frac{d^m}{dx^m}
  (x^2-1)^m \,.
\end{equation}
Furthermore, all the orthonormal polynomials will appear in their normalized version
\begin{align}
  &\tilde H_{ij}(\nu,J_1) \equiv
  \frac{1}{\sqrt{i!j!}} H_{ij}(\nu,J_1) \;, \label{eq:Hnorm} \\
  &\tilde L^{(\alpha)}_n (x) \equiv (-1)^n
  \sqrt{\frac{n!\,\Gamma(\alpha+1)}{\Gamma(n+\alpha+1)}}
  L^{(\alpha)}_n (x) \;, \label{eq:Lnorm} \\
  &\tilde P_m(x) \equiv \sqrt{2m+1} P_m(x) \;. \label{eq:Pnorm}
\end{align}
The normalized Laguerre $\tilde L_n^{(\alpha)}$, where $\alpha$ is related to the number of d.o.f. $k$ through 
$\alpha=k/2-1$, is defined such that the term with highest power always has a positive coefficient.
These polynomials satisfy the standard orthonormality conditions (see e.g. \cite{withers:2000})
\begin{align}
  & \int_{-\infty}^{+\infty} \!\!\! d\nu dJ_1\,
  \tilde{H}_{ij}(\nu,J_1)  \tilde{H}^\star_{lm}(\nu,J_1)
  {\cal N}(\nu,J_1) = \delta_{il} \delta_{jm} \;, \\
  &\int_0^\infty\!\! dx\,
  \tilde{L}_i^{(\alpha)}(x/2) \tilde{L}_j^{(\alpha)}(x/2)
  \chi^2_k(x) = \delta_{ij} \;, \\
  &\frac{1}{2} \int_{-1}^1\!\! dx \, 
  \tilde P_m(x) \tilde P_n(x)
  =\delta_{mn}\;.
\end{align}
In the last equality, the factor of $1/2$ arises from the fact that Legendre polynomials are orthonormal
relative to the weight $(1/2)\Theta_H(1-|x|)$. Note also that $\tilde L_n^{(\alpha)}(x/2)$ and $\tilde P_m(x)$ are 
their own dual.
Finally, we shall also take advantage of the formula
\begin{align}
  e^{X^2\!/2}(\nabla^{2}_{\!X})^n e^{-X^2\!/2} 
 & = (-2)^nn! L_n^{(k/2-1)}(X^2\!/2) \;,
\label{eq:Laguerreformula}
\end{align}
with $X^2=\sum_{i=1}^k X_i^2$ and $\nabla^{2}_{\!X}= \sum_{i=1}^k(\pd/\pd X_i)^2$. For $k=0$, it reduces to 
the well-known relation between Laguerre polynomials of weight $-1/2$ and univariate Hermite.

Since there is no mixing between operators of different rank, and since ${\cal N}(\vy)$ in \eqref{eq:multihermite} 
factorizes into ${\cal N}(\nu,J_1) {\cal N}(\bm{\eta}){\cal N}(\bar\zeta)$, so do the scalar, vector and tensor 
part of the orthonormal polynomials $\tilde O_\vn(\vw)$. 
The typical invariant contribution to \eqref{eq:multihermite} looks like 
\begin{equation}
  H_{ij}(\nu,J_1)\,
  \frac{(\nabla_\eta^{2})^q \mathcal{N}(\bm\eta)}{\mathcal{N}(\bm\eta)}\,
  \frac{(-1)^m D_2^lD_3^m\mathcal{N}(\bar\zeta)}{\mathcal{N}(\bar\zeta)} \;,
\end{equation}
with the bivariate Hermite polynomials defined in equation \eqref{eq:H} (these are not $SO(3)$ indices!). 
With aid of Eq.\eqref{eq:Laguerreformula} with $k=3$, it follows that terms with different $l$ are automatically 
orthogonal. 
Accounting for the normalization condition, the scalar and vector part of the orthonormal polynomials thus are
\begin{equation}
  \tilde H_{ij}(\nu,J_1)\, \tilde L_q^{(1/2)}(3\eta^2/2)\;,
\end{equation}
as expected.
Notice that the vector part will only appear with an even number of derivatives.

The tensor part is slightly more complicated, because it involves two different operators. The typical invariant 
term with $2l+3m$ derivatives arising from \eqref{eq:hermbias} will be
\begin{equation}
  \frac{(-1)^mD_2^lD_3^m\mathcal{N}(\bar\zeta)}{\mathcal{N}(\bar\zeta)} =
  e^{-D_2/2}J_2^lJ_3^m\,,
\end{equation}
which yields $(-2)^l l!L_l^{(3/2)}(5J_2/2)\equiv $ for $m=0$ , as it follows from setting $k=5$ in equation 
\eqref{eq:Laguerreformula}. However, terms with different values of $l$ and $m$ but the same total number of 
derivatives $2l+3m\geq6$ are in general not orthogonal. They must then be made orthogonal via the Gram-Schmidt 
procedure, which returns
\begin{align}
  \tilde F_{lm}(5 J_2,J_3)&= \sqrt{\frac{\Gamma(5/2)}{2^{3m}\Gamma(3m+5/2)}} \\
&\qquad \times  \tilde L_{l}^{(3m+3/2)}(s/2)s^{3m/2}\tilde P_m(x_3) \nonumber \;,
\end{align}
where $s=5J_2$, and $x_3=J_3/J_2^{3/2}$ as in Sec.\S\ref{sec:determinant}.
This procedure was first outlined by \cite{gay/pichon/pogosyan:2012} who, however, explicitly computed the two 
special cases $\tilde{F}_{l0}$ and $\tilde{F}_{01}$ solely. 
We use the notation $\tilde F_{lm}$ to emphasize the connection with their work, of which the above result is an 
important extension. Legendre polynomials arise because $x_3$ has a flat distribution $P(x_3)=\Theta_H(1-|x_3|)/2$.
The functions $\tilde F_{lm}$ are polynomials in $J_2$ and $J_3$. They satisfy
\begin{align}
&\int_0^\infty\!\!d(5J_2)\int_{-J_2^{3/2}}^{+J_2^{3/2}}\!\!dJ_3\, P(5J_2,J_3)
\tilde{F}_{lm}(5 J_2,J_3) \\ 
&\qquad \times \tilde{F}_{l'm'}(5 J_2,J_3) = \delta_{l l'}\delta_{m m'} \nonumber \;.
\end{align}
They are thus orthonormal with respect to the factorized weight 
$P(5J_2,J_3)=J_2^{-3/2}\chi_5^2(5 J_2)\Theta_H(J_2^{3/2}-|J_3|)/2$.

The invariant (normalized) polynomials thus are
\begin{equation}
\tilde{O}_{ijqlm}=\tilde H_{ij}(\nu,J_1) \tilde L_q^{(1/2)}(3\eta^2/2) \tilde F_{lm}(5 J_2,J_3)\;,
\end{equation}
and the corresponding weight is $\varpi=(1, 1, 2, 2, 3)$ for $\vw=(\nu,J_1,3\eta^2,5 J_2,J_3)$, respectively. 

\subsubsection{Perturbative expansion}

To write down the perturbative bias expansion, which generally assumes the form of Eq.(\ref{eq:generalizedbias}),
we need the bias coefficients and the dual polynomials. For BBKS peaks, the former are given by
\begin{align}
\label{eq:cijqlm}
  &\tilde{c}_{ijqlm} \\
&\equiv\frac{1}{\bnpk}
  \avg{\! \npk(\vw)\,\tilde{H}_{ij}(\nu,J_1)\tilde{L}_q^{(1/2)}\!(3\eta^2/2)
    \tilde{F}_{lm}(5 J_2,J_3)\!} \nonumber \;,
\end{align}
whereas the latter read
\begin{align}
\tilde{O}^\star_{ijqlm} &= \tilde{H}^\star_{ij}(\nu,J_1) \tilde L_q^{(1/2)}(3\eta^2/2) \\
& \qquad \times \tilde F_{lm}(5 J_2,J_3) \;, \nonumber
\end{align}
since $L_q$ and $F_{lm}$ are their own dual. The first few dual Hermites are
\begin{gather}
  H^\star_{n0}(\nu,J_1) = H_n(\nu)\;,
  \qquad H^\star_{0n}(\nu,J_1)= H_n(J_1) \;,
  \nonumber \\
  {H}^\star_{11}(\nu,J_1)
  = \nu J_1 - \gamma_1 \nonumber \;,\\
  {H}^\star_{21}(\nu,J_1)
  = \nu^2 J_1 - J_1 -2\gamma_1\nu \nonumber\;, \\
  {H}^\star_{12}(\nu,J_1)
  = \nu J_1^2 - \nu -2\gamma_1J_1 \nonumber \;,\\
  {H}^\star_{31}(\nu,J_1)
  = \nu^3 J_1 - 3\nu J_1 - 3\gamma_1\nu^2 +3\gamma_1 \nonumber \;,\\
  {H}^\star_{22}(\nu,J_1)
  = \nu^2 J_1^2 - \nu^2 - J_1^2 - 4\gamma_1\nu J_1 +1 +2\gamma_1^2 \;,
\end{gather}
where, like in standard Hermite polynomials, lower order terms appear with the right combinatorial coefficients 
to remove the 0-lag correlation functions from higher terms.
Recalling that $\tilde H^\star_{ij} = H^\star_{ij}/\sqrt{i!j!}$, the perturbative bias expansion of BBKS peaks is
\begin{align}
\label{eq:dpkBBKS}
  \dpk(\vx) &= \!\! \sum_\text{indices} \! \tilde{c}_{ijqlm}
  \tilde{H}^\star_{ij}\bigl(\nu(\vx),J_1(\vx)\bigr) \\
  &\quad \times
  \tilde{L}_j^{(1/2)}\!\left(3\eta^2(\vx)/2\right)
  \tilde{F}_{lm}\bigl(5 J_2(\vx),J_3(\vx)\bigr) \nonumber \\
  &= \tilde c_{10000}\, \tilde{H}_1(\nu(\vx))
  + \tilde{c}_{01000}\, \tilde{H}_1(J_1(\vx)) + \dots \nonumber
\end{align}
which, at second order, agrees with D13.

From Eqs.(\ref{eq:Hnorm}) - (\ref{eq:Pnorm}), the normalized bias coefficients are related to the usual bias factors 
$c_{ijqlm}$ through
\begin{align}
c_{ijqlm} &= (-1)^{q+l}
\sqrt{\frac{i!\, j!\, \Gamma(q+3/2)\,\Gamma(l+5/2)}{q!\,\Gamma(3/2)\,l!\,\Gamma(5/2)\,(2m+1)}} 
\nonumber \\
&\qquad \times \frac{\tilde{c}_{ijqlm}}{\sigma_0^i\sigma_1^{2q}\sigma_2^{j+2l+3m}} \;.
\label{eq:ctildec}
\end{align}
A factor of $(-1)^{q+l}$ is introduced to ensure that the terms with largest power of $\eta^2$ and $J_2$ in the 
perturbative bias expansion always have positive sign. 
The $c_{ijqlm}$ are the coefficients of the series expansion, had it been written in terms of the fields $\delta$, 
$(\grad\delta)^2$ etc. and without the normalisation as is conventional in large scale structure. 
Namely, $\dpk(\vx)$ can also be written
\begin{align}
\label{eq:olddpk}
&\dpk(\vx) = \bigg. c_{10000}\delta_R(\vx) - c_{01000} \nabla^2\delta_R(\vx) \\
& \quad + \bigg.\frac{c_{20000}}{2}\Big(\delta_R^2(\vx)-\sigma_0^2\Big) 
- c_{11000} \Big(\delta_R(\vx) \nabla^2\delta_R(\vx)-\sigma_1^2\Big) \nonumber \\
& \quad + \bigg. \frac{c_{02000}}{2}\Big[\big(\nabla^2\delta_R\big)^2(\vx)-\sigma_2^2\Big]
+c_{00100} \Big[\big(\grad\delta_R\big)^2\!(\vx)-\sigma_1^2\Big] \nonumber \\
& \quad+ c_{00010}\bigg[\frac{3}{2}\bigg(\partial_{ij}\delta_R-\frac{1}{3}\delta_{ij}\nabla^2\delta_R\bigg)^2\!(\vx)
-\sigma_2^2\bigg]  +\dots \nonumber
\end{align}
We have $c_{ij000}\equiv b_{ij}$, $c_{00q00}\equiv \chi_q$ and $c_{000lm}\equiv \omega_{lm}$, even though 
$c_{ijqlm}\neq b_{ij}\,\chi_q\,\omega_{lm}$ in general.
Since the weight are $\varpi=(1,1,2,2,3)$, the terms of order $n$ are those which satisfy $i+j+2q+2l+3m=n$. 
Note also that one could equally work with the independent variables $(J_1,z,...)$  (rather than $(\nu,J_1)$), 
such that univariate (rather than bivariate) Hermite polynomials appear in the expansion. 

\subsubsection{Relation to peak-background split}

All the bias factors $c_{ijqlm}$ can be obtained from a peak-background split, as shown in D13. For instance, the bias
parameters $c_{00q00}\equiv \chi_q$ associated to the chi-square variable $3\eta^2$ can be seen as the response of the
peak number density $\bnpk$ to a long mode $\sigma_1^2\eta_\text{\tiny L}^2$ ($=(\grad\delta)_\text{\tiny L}^2$), 
with $\veta_\text{\tiny L}=(\eta_{\text{\tiny L}1},\eta_{\text{\tiny L}2},\eta_{\text{\tiny L}3})$,
\begin{equation}
c_{00q00}= \frac{1}{\bnpk}\frac{\partial^{2q}\bnpk}{\partial(\sigma_1\!\eta_\text{\tiny L})^{2q}}\equiv \chi_q \;.
\end{equation}
Similarly, the bias factors $c_{000lm}$ can be interpreted with the peak-background split as
\begin{equation}
c_{000lm}= (-1)^m\frac{1}{\bnpk}\frac{\partial^{l+m}\bnpk}
{\partial(\sigma_2^2 J_{2\text{\tiny L}})^l\partial(\sigma_2^3 J_{3\text{\tiny L}})^m}
\equiv \omega_{lm}\;,
\end{equation}
where $\sigma_2^2 J_{2\text{\tiny L}}\sim {\rm tr}(\bar\zeta_{ij}^2)_\text{\tiny L}$ and 
$\sigma_2^3 J_{3\text{\tiny L}}\sim {\rm tr}(\bar\zeta_{ij}^3)_\text{\tiny L}$ are 
long-wavelength perturbations to the quadratic and cubic traces $J_2$ and $J_3$, respectively.

It is not difficult to see that $\big\langle \npk\tilde{F}_{l0}\big\rangle$ is proportional to the bias factors 
$\omega_{l0}$ (which is defined as $\chi_{0l}$ in D13). 
After some algebra, we indeed find that Eq.(\ref{eq:dpkBBKS}) truncated at second-order exactly reproduces the peak 
correlation $\xpk(r)$ computed by \cite{desjacques/crocce/etal:2010} (as was already noted in D13).
For example, the second-order term involving $\eta^2$ simplifies to
\begin{align}
\big\langle \tilde{L}_1^{(1/2)}\!(3\eta^2&/2)\big\rangle \tilde{L}_1^{(1/2)}\!(3\eta^2(\vx)/2) \\
&= \frac{2}{3} \big\langle L_1^{(1/2)}\!(3\eta^2/2)
\big\rangle\left(\frac{3}{2}-\frac{3}{2}\eta^2(\vx)\right) \nonumber \\
&= \sigma_1^2 \chi_1 \left(\eta^2(\vx)-1\right), \nonumber
\end{align}
where we have used $\chi_1=-3/(2\sigma_1^2)$. Thus, it contributes
\begin{align}
\label{eq:term2ndeta}
\sigma_1^4\chi_1^2&\Big\langle\big(\eta^2(\vx_1)-1)(\eta^2(\vx_2)-1\big)\Big\rangle \\
&= \sigma_1^4\chi_1^2\Big(\big\langle\eta^2(\vx_1)\eta^2(\vx_2)\big\rangle-1\Big) \nonumber \\
&= 2\sigma_1^4\chi_1^2 \avg{\eta_i(\vx_1)\eta_j(\vx_2)}\avg{\eta_i(\vx_1)\eta_j(\vx_2)} \notag \\
&= \frac{3}{2\sigma_1^4}\Bigl[\bigl(\xi_0^{(1)}\!(r)\bigr)^2+2\bigl(\xi_2^{(1)}\!(r)\bigr)^2\Bigr] \;,
\nonumber 
\end{align}
to the 2-point peak correlation function $\xpk(r)$, with $r=|\vx_1-\vx_2|$.
The functions $\xi_\ell^{(n)}\!(r)$ are quantities analogous to $\sigma_n^2$ but defined for a finite 
separation $r$,
\begin{equation}
\xi_\ell^{(n)}\!(r)= \frac{1}{2\pi^2}\int_0^\infty\!\! dk\,
k^{2(n+1)}\,\tilde{W}_R^2(k)\,  P_L(k)\; j_\ell(kr)\;,
\label{xielln}
\end{equation}
where $j_\ell(x)$ are spherical Bessel functions. This also illustrates the point that the bias factors 
are renormalized since the fields appear in the argument of orthogonal polynomials.

Interestingly, no term with only one $\avg{\eta_i(\vx_1)\eta_j(\vx_2)}$ will appear in the peak-peak correlation 
function. 
While such a term would be allowed for the Gaussian variables at finite separation, due to the breaking of $SO(3)$ 
invariance by the separation vector $\vx_2-\vx_1$, odd powers of $\boldsymbol{\eta}$ do not appear in the expansion 
of $\dpk(\vx)$. 
Similarly, no exposed indices are allowed, nor terms in which the index contractions do not respect the 
$SO(3)\times SO(3)$ invariance.
This is another effect of the larger symmetry of the peak model.

Note that, in practice, the various Lagrangian bias coefficients can be measured directly from the 
simulations by projection, that is, by cross-correlating the Lagrangian halos with the appropriate 
combination of orthogonal polynomials 
\citep[see][for such measurements]{musso/paranjape/sheth:2012,paranjape/sheth/desjacques:2013,biagetti/chan/etal:2014}.

\section{Non-Gaussian bias from a primordial trispectrum} 
\label{sec:PNG}

As an illustration, we will apply our result to the calculation of the non-Gaussian bias induced by a local 
primordial trispectrum. This shape arises in cubic PNG, in which the spatial curvature $\Phi$ is given by 
$\Phi(\vx)=\phi(\vx)+\gnl\phi^3(\vx)$ ($\phi$ is the Gaussian part).

We will consider BBKS peaks and evaluate $\left\langle\dpk(\mb{x}_1) \dpk(\mb{x}_2)\right\rangle$ from the 
perturbative bias expansion Eq.(\ref{eq:dpkBBKS}) at $4^{\rm th}$ order in the fields. 
We will then show that our result agrees with a peak-background split prediction. 

\subsection{Effective bias expansion with PNG}

As shown in \cite{desjacques/gong/riotto:2013}, the computation of the effect of PNG on clustering statistics 
of biased tracers proceeds exactly like in the Gaussian case, except for the fact that:
\begin{itemize}
\item The 1-point PDF $P(\vy)$ is non-Gaussian. This generates {\it scale-independent} corrections to
the biased factors $c_{ijqlm}$. 
However, we will neglect them here because this is not the dominant effect at large scales.
\item The fields $\nu(\vx)$, $J_1(\vx)$, $\eta^2(\vx)$ etc. are non-Gaussian. 
This induces {\it scale-dependent} corrections to the 2-point correlation $\xpk(r)$ which, 
for a local primordial trispectrum, blow up in the limit $k\to 0$. These are the focus of this Section.
\end{itemize}
In general, the contributions to the non-Gaussian bias arising from a primordial trispectrum can arise 
either from ``(2-2) correlators'', which correspond to ensemble averages of a product of two $2^{\rm nd}$ 
order quantities at two distinct locations, or from ``(1-3) correlators'', which designate products 
of a $1^{\rm st}$ order term (i.e $\nu(\vx)$ or $J_1(\vx)$) with a $3^{\rm rd}$ order term. 

\subsubsection{(2-2) correlators} 
\label{subsec:2ndorder}

We begin with the calculation of the connected 4-point functions of (2-2) type. Previous work has 
shown that they do not contribute to the $k$-dependent non-Gaussian bias. 
Nevertheless, we will briefly discuss them for sake of completeness. All these contributions will
be of the form 
\begin{equation}
\big\langle \nu^2(\vx_1) \nu^2(\vx_2) \big\rangle \;,
\end{equation}
e.g. $\big\langle \nu^2(\vx_1) \nu(\vx_2) J_1(\vx_2)\big\rangle$, 
$\big\langle \nu^2(\vx_1)J_2(\vx_2)\big\rangle$. 
Each of these 4-point correlators can be schematically decomposed as
\begin{align}
\label{eq:corrschem}
\big\langle (\cdot)_1 (\cdot)_1 (\cdot)_2 (\cdot)_2 \big\rangle &= 
\big\langle (\cdot)_1 (\cdot)_1 \big\rangle_c \big\langle (\cdot)_2 (\cdot)_2 
\big\rangle_{\rm c} 
\\ & \quad 
+ 2 \cdot \big\langle (\cdot)_1 (\cdot)_2 \big\rangle_c \big\langle (\cdot)_1 (\cdot)_2 
\big\rangle_{\rm c} 
\nonumber \\ & \quad 
+ \big\langle (\cdot)_1 (\cdot)_1 (\cdot)_2 (\cdot)_2 \big\rangle_{\rm c} \nonumber \;,
\end{align}
where each of the $(\cdot)$ schematically represents one of the variable entering the effective bias expansion, ``c'' stands for connected, and the subscripts ``1'' and ``2'' indicate quantities evaluated at spatial
position $\vx_1$ and $\vx_2$, respectively. The connected 4-point correlators all have the same structure and 
can be generically written as 
\begin{align}
\label{eq:connecschem}
\big\langle (\cdot)_1 (\cdot)_1 (\cdot)_2 (\cdot)_2 &\big\rangle_{\rm c} =
g_1\!\left(\sigma_0,\sigma_1, \sigma_2\right) 
\Bigg\{\prod^4_{i=1} \int\!\! \frac{d^3 k_i}{(2 \pi)^3}\Bigg\} \\
& \qquad \times
\left\langle \delta_R(\mb{k}_1) \delta_R(\mb{k}_2) \delta_R(\mb{k}_3) \delta_R(\mb{k}_4) \right\rangle_{\rm c}
\nonumber \\ &\qquad 
\times g_2\!\left(\mb{k}_1, \mb{k}_2, \mb{k}_3, \mb{k}_4\right) \,
e^{i \mb{k}_{12}\cdot \mb{x}_1 + i \mb{k}_{34}\cdot \mb{x}_2} \;,\nonumber
\end{align}
where we used again the notation, $\vk_{i \dots n}=\vk_i+ \dots +\vk_n$, while $g_1$ and $g_2$ are general 
functions of the spectral moments $\sigma_i$ and wavemodes $\vk_i$, respectively. Using
\begin{align}
\label{eq:TSdta}
\big\langle \delta_R(\mathbf{k}_1)& \delta_R(\mathbf{k}_2) \delta_R(\mathbf{k}_3) \delta_R(\mathbf{k}_4) 
\big\rangle_{\rm c} \\ 
&= (2 \pi)^3 \, T_R(k_1, k_2, k_3, k_4)\,
\delta_{\rm D} \left(\mathbf{k}_1 + \mathbf{k}_2 + \mathbf{k}_3 + \mathbf{k}_4\right) \nonumber \;,
\end{align}
with $T_R$ being the trispectrum of the linear, smoothed density field $\delta_R$, we get
\begin{align}
\big\langle (\cdot)_1 & (\cdot)_1 (\cdot)_2 (\cdot)_2 \big\rangle_{\rm c} = 
g_1\!\left(\sigma_0,\sigma_1, \sigma_2\right) \label{eq:connecschemT} \\
&\qquad \times \prod^3_{i=1} \int\!\!\frac{d^3 k_i}{(2 \pi)^3}\, e^{i \mathbf{k}_{12} \mathbf{r}} 
g_2\left(\mb{k}_1, \mb{k}_2, \mb{k}_3, - \mb{k}_{123}\right) \nonumber \\
&\qquad \times
T_R\left(k_1, k_2, k_3, k_{123}\right) \;, \nonumber
\end{align}
where $\mathbf{r} \equiv \mb{x}_1 - \mb{x}_2$ is the separation vector, and $k_{ijq}=|\vk_{ijq}|$. On performing the Fourier 
transform of \eqref{eq:connecschemT}, substituting the Fourier representation of the Dirac delta, 
$\delta_{\rm D} (\mb{k}) = \int{d^3\mb{r} e^{-i \mb{k}\cdot\mb{r}}}$, integrating over $\mb{k}_1$ 
and taking the limit $k\rightarrow 0$, we arrive at 
\begin{align}
\label{eq:FT4ptFinal}
\int d^3\mathbf{r}\big\langle (\cdot)_1 & (\cdot)_1 (\cdot)_2 (\cdot)_2 \big\rangle_{\rm c} 
e^{-i\mathbf{k}\mathbf{r}} \stackrel{k\rightarrow0}{=} 
g_1\!\left(\sigma_0,\sigma_1, \sigma_2\right)  \\
&\qquad\times
\iint \frac{d^3 \mathbf{k}_1}{(2 \pi)^3} \frac{d^3 \mathbf{k}_2}{(2 \pi)^3}\, 
g_2\!\left(-\vk_1,\vk_1,\vk_2,-\vk_2\right) \nonumber \\
&\qquad \times T_R\left(k_1, k_1, k_2, k_2\right) \nonumber \;.
\end{align}
Even though we have not specified the exact form of $g_1$, $g_2$ or $T_R$, it is obvious that there is 
no $k$-dependence in this expression. 
Therefore, it can be ignored in the limit $k\rightarrow 0$ even though it can be significantly larger 
than the typical shot noise contribution to the halo power spectrum \cite{desjacques/seljak:2010}.

\subsubsection{(1-3) correlators} 
\label{subsec:4thorder}

Taking into account the factors of two arising from the exchange of $\mb{x}_1$ and $\mb{x}_2$, the 
leading non-Gaussian contribution to the 2-point correlation of BBKS peaks in the limit $k\to 0$ is 
twice the following sum of $(1 - 3)$ correlators,
\begin{widetext}
\begin{align}
&\big\langle \tilde{H}_{10}(\nu,J_1)\big\rangle
\sum_{ijqlm}
\tilde c_{ijqlm}
\bigg\langle \tilde{H}^\star_{10}(\nu(\vx_1),J_1(\vx_1))
\tilde{H}^\star_{ij}(\nu(\vx_2),J_1(\vx_2))
\tilde{L}_q^{(1/2)}\!\!\left(\frac{3}{2}\eta^2(\vx_2)\right)
\tilde{F}_{lm}\!\!\left(5J_2(\vx_2),J_3(\vx_2)\right)\bigg\rangle  
 \\
&\qquad +\big\langle \tilde{H}_{01}(\nu,J_1)\big\rangle
\sum_{ijqlm} \tilde c_{ijqlm}
\bigg\langle \tilde{H}^\star_{01}(\nu(\vx_1),J_1(\vx_1))\tilde{H}^\star_{ij}(\nu(\vx_2),J_1(\vx_2))
\tilde{L}_q^{(1/2)}\!\!\left(\frac{3}{2}\eta^2(\vx_2)\right)
\tilde{F}_{lm}\!\!\left(5J_2(\vx_2),J_3(\vx_2)\right)\bigg\rangle \;. 
\nonumber
\end{align}
The sum runs over the indices $\{i,j,q,l,m\}$ subject to the constraint $i+j+2q+2l+3m=3$. 
For instance, the term 
$\big\langle \tilde{H}^\star_{01}(\nu(\vx_1),J_1(\vx_1))\tilde{H}^\star_{30}(\nu(\vx_2),J_1(\vx_2))\big\rangle$ 
reduces to
\begin{align}
\Big\langle \tilde{H}_1(J_1(\vx_1))& \tilde{H}_3(\nu(\vx_2))\Big\rangle \\ 
&= \frac{1}{\sqrt{3!}}
\Bigl(\big\langle J_1(\vx_1)\nu^3(\vx_2)\big\rangle - 3\big\langle J_1(\vx_1)\nu(\vx_2)\big\rangle\Bigr) 
\nonumber \\
&= \frac{1}{\sqrt{3!}} \big\langle J_1(\vx_1)\nu^3(\vx_2)\big\rangle_c \nonumber \;,
\end{align}
while the term 
$\big\langle \tilde{H}^\star_{10}(\nu(\vx_1),J_1(\vx_1))\tilde{H}^\star_{10}(\nu(\vx_2),J_1(\vx_2)) 
\tilde{L}_1^{(1/2)}\!(3\eta^2(\vx_2)/2)\big\rangle$ 
simplifies to
\begin{align}
\Big\langle \tilde{H}_1(&\nu(\vx_1))\tilde{H}_1(\nu(\vx_2))\tilde{L}_1^{(1/2)}\!(3\eta^2(\vx_2)/2)\Big\rangle \\
& =
\sqrt{\frac{2}{3}}\bigg\langle\nu(\vx_1)\nu(\vx_2)\left(\frac{3}{2}-\frac{3}{2}\eta^2(\vx_2)\right)\bigg\rangle 
\nonumber \\
&= \sqrt{\frac{3}{2}}
\biggl(\big\langle\nu(\vx_1)\nu(\vx_2)\big\rangle-\big\langle\nu(\vx_1)\nu(\vx_2)\big\rangle
\big\langle\eta^2(\vx_2)\big\rangle-\big\langle\nu(\vx_1)\nu(\vx_2)\eta^2(\vx_2)\big\rangle_c\biggr) \nonumber \\
&= -\sqrt{\frac{3}{2}}\big\langle\nu(\vx_1)\nu(\vx_2)\eta^2(\vx_2)\big\rangle_c \nonumber \;.
\end{align}
This illustrates the point that correlators of terms in the perturbative peak bias expansion with 
different values of $n=i+j+2q+2l+3m$ vanish unless the initial conditions are non-Gaussian.

After further simplifications, the dominant correction to the peak power spectrum in the low-$k$
limit is
\begin{equation}
\Delta P_\text{pk}(k) = \int\!\! d^3r\, \Delta \xi_{\rm pk}(r)\, e^{-i \mb{k} \cdot \mb{r}} \;, 
\label{eq:DPS4}
\end{equation}
where $\Delta \xi_\text{pk}(r)$ is the sum of all correlators of $(1-3)$ type,
\begin{align}
\label{eq:DXi4}
\Delta\xi_\text{pk}(r) &= 2 \big\langle \tilde{H}_{10}(\nu,J_1)\big\rangle
\bigg\{\frac{\tilde c_{30000}}{\sqrt{6}}\big\langle\nu(\vx_1)\nu^3(\vx_2)\big\rangle_c +
\frac{\tilde c_{21000}}{\sqrt{2}}\big\langle\nu(\vx_1)\nu^2(\vx_2)J_1(\vx_2)\big\rangle_c  \\
&\quad +\frac{\tilde c_{12000}}{\sqrt{2}}\big\langle\nu(\vx_1)\nu(\vx_2) J_1^2(\vx_2)\big\rangle_c
+\frac{\tilde c_{03000}}{\sqrt{6}}\big\langle\nu(\vx_1) J_1^3(\vx_2)\big\rangle_c 
\nonumber \\
&\quad-\sqrt{\frac{3}{2}} \biggl[\tilde c_{10100}
\big\langle\nu(\vx_1) \nu(\vx_2)\eta^2(\vx_2)\big\rangle_c + \tilde c_{01100}
\big\langle\nu(\vx_1) J_1(\vx_2)\eta^2(\vx_2)\big\rangle_c \biggr]
\nonumber \\
&\quad -\sqrt{\frac{5}{2}} \biggl[\tilde c_{10010}\big\langle\nu(\vx_1) \nu(\vx_2)J_2(\vx_2)\big\rangle_c 
+ \tilde c_{01010} \big\langle\nu(\vx_1) J_1(\vx_2)J_2(\vx_2)\big\rangle_c \biggr]
\nonumber \\
&\quad+\frac{5}{\sqrt{21}} \tilde c_{00001}\big\langle\nu(\vx_1) J_3(\vx_2)\big\rangle_c \bigg\} 
+ \nu(\vx_1) \leftrightarrow J_1(\vx_1) \nonumber \;,
\end{align} 
\end{widetext}
where $\nu(\vx_1)\leftrightarrow J_1(\vx_1)$ indicates that similar terms appear, with all occurrences of $\nu(\vx_1)$
replaced by $J_1(\vx_1)$. Our definition of the normalized bias factors $\tilde c_{ijqlm}$ brings along additional 
factors of $1/\sqrt{i!}$ etc. when $\Delta\xi_\text{pk}(r)$ is expressed in terms of the usual bias factors
$c_{ijqlm}$. We thus have 18 correlators, each of which is multiplied by two in order to account for the exchange of 
$\vx_1$ and $\vx_2$.

\subsubsection{Skewness}

We shall now calculate the Fourier transform of the above connected correlators in the limit $k\rightarrow 0$. 
In term of the power spectrum $P_\phi(k)$ of the Gaussian potential $\phi$, the primordial trispectrum takes the local 
shape 
\begin{equation}
T_{\phi}^{\gnl}\!(k_1, k_2, k_3, k_4) = 
6 \gnl \Bigl[P_{\phi} (k_1)P_{\phi} (k_2)P_{\phi} (k_3) + \mbox{3 cyc.} \Bigr] \;.
\label{eq:TS}
\end{equation}
The calculation is rather straightforward and we schematically obtain the following result:
\begin{align}
\label{eq:schematic}
\int\!\! d^3r \,\left\langle X_1 X_2 \right\rangle_c e^{-i \mb{k}\cdot\mb{r}} 
&\stackrel{k\rightarrow0}{=} 3 \gnl P_{\phi} (k) \mathcal{M}_R(k) {\cal X}_1(k) \\
&\qquad \times S^{(X_2)}_3\!(M)\,  g(\sigma_0,\sigma_1,\sigma_2) \nonumber \;.
\end{align}
Here, ${\cal M}_R(k)\equiv {\cal M}(k)\tilde{W}_R(k)$ is the transfer function between the potential $\phi$ and 
the linear, smoothed density perturbation $\delta_R$. 
Furthermore, $X_1(k)\equiv {\cal X}_1(k)\delta_R(k)$ where $X_1(k)$ is the Fourier transform of $X_1=X(\vx_1)$, 
and $g(\sigma_0,\sigma_1,\sigma_2)$ is a function of the spectral moments.
Finally, $S_3^{(X_2)}\!(M)$ is a generalized skewness which depends on the details of the fields at position 
$\vx_2$, i.e. $X_2=X(\vx_2)$. 
Since there is no absolute necessity to go into much details, we loosely define $S_3^{(X_2)}(M)$ as
\begin{align}
\label{eq:skewness}
\sigma^4_0 &S_3^{(X_2)}\!(M)= 
2\iint \frac{d^3k_1}{(2\pi)^3} \frac{d^3k_2}{(2\pi)^3} {\cal X}_2(\vk_1,\vk_2)\mathcal{M}_R(k_1) 
\nonumber \\
&\times\mathcal{M}_R(k_2) \mathcal{M}_R(k_{12})
\Big[P_{\phi} (k_1) P_{\phi} (k_2) +\mbox{2 cyc.}\Big]  \;.
\end{align}
For illustration, in the simplest case of $\big\langle\nu(\vx_1)\nu^3(\vx_2)\big\rangle_c$, 
we find $g=\sigma_0$, ${\cal X}_1(k)=1/\sigma_0$ and ${\cal X}_2=1$ whereas, for 
$\big\langle J_1(\vx_1)\nu(\vx_2)\eta^2(\vx_2)\big\rangle_c$, we have $g=\sigma_0^3/\sigma_1^2$,
${\cal X}_1(k)=k^2/\sigma_2$ and ${\cal X}_2=-\vk_1\cdot\vk_2$.

Fourier space correlators with $\nu(\vx_1)$ being replaced by $J_1(\vx_1)$ are identical to those 
involving $\nu(\vx_1)$ except for a multiplicative factor of $k^2/\sigma_2$ instead of $1/\sigma_0$. 
Hence, they become negligible in the limit $k \rightarrow 0$. 

\subsection{Non-gaussian correction to the first bias factor} 
\label{subsec:NGcorrbias}

On summing all the contributions from the (1-3) correlators, the non-Gaussian contribution to the peak
power spectrum in the limit $k\to 0$ is
\begin{align}
\label{eq:DPkFinal}
\Delta P_{\rm pk}(k)&= 6\gnl P_{\phi} (k) \mathcal{M}_R(k) \\
&\quad\times \bigg(
\frac{1}{\sigma_0}\big\langle H_{10}(\nu_1,J_1)\big\rangle +\frac{1}{\sigma_2}
\big\langle H_{01}(\nu_1,J_1)\big\rangle k^2\bigg) \nonumber \\
&\quad\times\sigma_0^4\bigg\{\frac{\tilde c_{30000}}{\sqrt{6}\sigma_0^3}S_3^{(\nu^3)}
+\frac{\tilde c_{21000}}{\sqrt{2}\sigma_0^2\sigma_2}S_3^{(J_1\nu^2)} \nonumber \\
&\qquad +\frac{\tilde c_{12000}}{\sqrt{2}\sigma_0\sigma_2^2}S_3^{(\nu J_1^2)} 
+ \frac{\tilde c_{03000}}{\sqrt{6}\sigma_2^3}S^{(J_1^3)} \nonumber \\
&\qquad -\sqrt{\frac{3}{2}}\biggl[\frac{\tilde c_{10100}}{\sigma_0\sigma_1^2}S_3^{(\nu\eta^2)}
+\frac{\tilde c_{01100}}{\sigma_1^2\sigma_2}S_3^{(J_1\eta^2)} \biggr] \nonumber \\
&\qquad -\sqrt{\frac{5}{2}}\biggl[\frac{\tilde c_{10010}}{\sigma_0\sigma_2^2}S_3^{(\nu J_2)} 
+\frac{\tilde c_{01010}}{\sigma_2^3}S_3^{(J_1 J_2)}\biggr] \nonumber \\
&\qquad +\frac{5}{\sqrt{21}}\frac{\tilde c_{00001}}{\sigma_2^3}S_3^{(J_3)}
\bigg\}
\nonumber \;.
\end{align}
Note that $\tilde{H}_{10}=H_{10}$. The first term in parentheses in the right-hand side, together with the filter 
$\tilde{W}_R(k)$ in $\mathcal{M}_R(k)$, simplifies to 
\begin{align}
\bigg[\frac{1}{\sigma_0}&\big\langle H_{10}(\nu_1,J_1)\big\rangle +\frac{1}{\sigma_2}
\big\langle H_{01}(\nu_1,J_1)\big\rangle k^2\bigg]\tilde{W}_R(k) \\
&= \bigg[\frac{1}{\sigma_0}\left(\frac{\nu_c-\gamma_1\bar J_1}{1-\gamma_1^2}\right)
+\frac{1}{\sigma_2}\left(\frac{\bar J_1-\gamma_1\nu_c}{1-\gamma_1^2}\right)k^2\bigg]\tilde{W}_R(k) 
\nonumber \\
&= \Big(b_{10}+b_{01} k^2\Big) \tilde{W}_R(k) \nonumber \\
&\equiv c_1^L(k) \nonumber\;,
\end{align}
with $\bar{J}_1 =\frac{1}{\bnpk}\big\langle\npk J_1\big\rangle$. This last expression is nothing else than 
the linear Lagrangian peak bias. 

Using Eq.(\ref{eq:ctildec}) to replace $\tilde c_{ijqlm}$ by the usual bias parameters $c_{ijqlm}$, the 
non-Gaussian correction to the linear bias is given by
\begin{align}
\Delta c_1^L(k) &= 3\gnl \mathcal{M}^{-1} (k) \sigma_0^4
\bigg\{\frac{1}{6}c_{30000}S_3^{(\nu^3)} \nonumber \\
& +\frac{1}{2} c_{21000} S_3^{(J_1\nu^2)} +\frac{1}{2}c_{12000}S_3^{(\nu J_1^2)} 
+\frac{1}{6} c_{03000}S_3^{(J_1^3)} \nonumber \\
& + c_{10100}S_3^{(\nu\eta^2)} + c_{01100}S_3^{(J_1\eta^2)} + c_{10010}S_3^{(\nu J_2)} \nonumber \\
& + c_{01010}S_3^{(J_1J_2)} +\frac{5}{\sqrt{7}} c_{00001}S_3^{(J_3)}\bigg\} \;,
\label{eq:DB1}
\end{align}
which follows from approximating the peak power spectrum as 
$P_{\rm pk} (k)\approx \big(c_1^L(k)\big)^2 P_L(k) + 2 c_1^L(k) \Delta c_1^L(k) P_L(k)$, 
and identifying the second term with $\Delta P_{\rm pk} (k)$.
Note that $c_{00001}=(5/3\sqrt{7}) J_3/\sigma_2^3$. Furthermore, $c_{30000}=b_3$  in the standard local bias. 
Therefore, if this term where the sole contribution to the non-Gaussian bias, then the non-Gaussian bias amplitude 
would be proportional to $(1/2)\gnl \sigma_0^4 S^{(\nu^3)} b_3$, in agreement with previous local bias calculations
(see \cite{desjacques/seljak:2010}).

Finally, as already noted, in the presence of PNG, the distribution $P(\mb{y})$ is not a multivariate Normal anymore, such that the 
bias factors $\tilde c_{ijqlm}$ are not precisely equal to their Gaussian counterparts. 
However, these corrections are scale-independent and, therefore, can be neglected in the limit $k\rightarrow 0$ 
\cite{afshordi/tolley:2008,desjacques/seljak/iliev:2009}.

\subsection{Relation to peak-background split}
\label{subsec:NGcorrbiasPBS}

In Appendix \S\ref{sec:PBSa}, we calculate the response of the peak number density $\bnpk$ to a 
long-wavelength perturbation in the potential. 
A comparison between the perturbative expression Eq.(\ref{eq:DB1}) and the peak-background split
result Eq.(\ref{eq:deltaPNG}) reveals that $\Delta c_1^L(k)$ can also be expressed as
\begin{align}
\label{eq:pbscubic}
\Delta c_1^L(k) &= \frac{1}{\bar{n}_{\rm pk}}
\frac{\partial\delta\bar{n}_{\rm pk}^\text{\tiny L}}{\partial\delta_\text{\tiny L}}
\Bigg\lvert_{\delta_\text{\tiny L}=0} \\\
&= 3\gnl {\cal M}^{-1}(k)\frac{\delta\bar{n}_{\rm pk}^{\fnl=1}}{\bnpk} \nonumber \;,
\end{align}
in agreement with the findings of \cite{desjacques/jeong/schmidt:2011,smith/ferraro/loverde:2012}. 
This demonstrates that our effective bias expansion yields the correct non-Gaussian bias in the presence 
of a local primordial trispectrum.

We can rewrite Eq.(\ref{eq:DB1}) in a more compact form upon substituting the explicit expression 
(\ref{eq:skewness}) for $\sigma_0^4S_3^{(X_2)}$. We find
\begin{align}
\Delta c_1^L(k) &= 3 \gnl \mathcal{M}^{-1}(k) \iint \frac{d^3k_1}{(2\pi)^3} \frac{d^3k_2}{(2\pi)^3}\,
c_3^L(\vk_1,\vk_2,\vk_3) \nonumber \\
&\qquad \times {\cal M}(k_1){\cal M}(k_2){\cal M}(k_3) P_\phi(k_1)P_\phi(k_2) \nonumber \\
&\qquad \Big. \times\delta_D(\vk_1+\vk_2+\vk_3) \;,
\label{eq:DB2} 
\end{align}
where 
\begin{align}
c_3^L(&\vk_1,\vk_2,\vk_3) = c_{30000} + c_{21000} \Big(k_1^2 + \mbox{2 cyc.}\Big) \\
& + c_{12000} \Big(k_1^2 k_2^2 + \mbox{2 cyc.}\Big) + c_{03000} k_1^2 k_2^2 k_3^2 
-2 c_{10100} \nonumber \\
& \quad \times \Big(\vk_2\cdot\vk_3 + \mbox{2 cyc.} \Big) 
-2 c_{01100}\Big[k_1^2 \big(\vk_2\cdot\vk_3) + \mbox{2 cyc}\Big] \nonumber \\
& + c_{10010} \Big[\big(3(\vk_1\cdot\vk_2)^2 - k_1^2 k_2^2\big) + \mbox{2 cyc.}\Big] \nonumber \\
& + c_{01010} \Big[k_1^2 \big(3(\vk_2\cdot\vk_3)^2 - k_2^2 k_3^2\big) + \mbox{2 cyc.}\Big] \nonumber \\
&- \frac{5\cdot 3^3}{\sqrt{7}} c_{00001} 
\bigg[\big(\vk_1\cdot\vk_2\big)\big(\vk_2\cdot\vk_3\big)\big(\vk_3\cdot\vk_1\big) \nonumber \\
&\quad -\frac{1}{3}\Big[\left(\vk_1\cdot\vk_2\right)k_3^2+\mbox{2 cyc.}\Big]+\frac{2}{9}k_1^2 k_2^2 k_3^2\bigg]\;,
\end{align}
is the third-order Lagrangian bias of BBKS peaks. 
Eq.(\ref{eq:DB2}) agrees with the iPT calculation of \cite{yokoyama/matsubara:2013}. 
Eqs.~(\ref{eq:pbscubic}) and (\ref{eq:DB2}) imply that, in the weak PNG limit considered here, the third-order 
biases satisfy a consistency relation, i.e. the integral in the rhs of Eq.(\ref{eq:DB2}) must equal 
$\delta\bar{n}_{\rm pk}^{\fnl=1}/\bnpk$.
Similar relations arise for local, quadratic PNG \cite{desjacques/gong/riotto:2013,biagetti/desjacques:2015}
and, presumably, for other types of PNG.

To conclude, let us stress that, although we have assumed that halo collapse proceeds according through the 
spherical collapse approximation, so that $\nu(\vx)=\nu_c$ (i.e. our collapse barrier is flat and deterministic), 
the above equality would also hold for a moving and stochastic barrier since the leading order non-Gaussian 
contribution enters through the skewness.

\section{Discussion and conclusion}
\label{sec:conclusion}

We have developed a general methodology that allows us to write down perturbative bias expansions in 
Lagrangian space whenever the biased tracers can be represented by a set of constraints in the initial
conditions. Our approach, which generalizes the work of \cite{szalay:1988} and \cite{desjacques:2013}, 
is generic and can thus be applied to tracers others than the linear density 
peaks or thresholded regions usually considered in the literature. Potential applications in addition
to galaxies include e.g. voids and the skeleton of the cosmic web.
The bias expansion can be used to evaluate any $N$-point connected Lagrangian correlation function 
perturbatively. Its development in orthogonal polynomials ensures that the series coefficients, the 
so-called Lagrangian bias parameters, are renormalized. The latter are ensemble averages over the 
subset of space that satisfies the constraints and, therefore, can be calculated at any order once the
constraints are known.

We have clarified the dependence of peak clustering on invariants such as the ``determinant'' $J_3$,
which has allowed us to derive the well-known BBKS relation for $\bnpk$ in a simple way.
In addition, we have shown that these variables behave like dynamical angles, and generate bias terms
which are proportional to Legendre polynomials $P_m$. 
Note, however, that the corresponding bias coefficients do not generally take the simple form
\begin{equation}
\frac{1}{\bnpk}\int\!\!d\vw\,\cpk(\vw)\, P_m(J_3/J_2^{3/2}) P(\vw)\:,
\end{equation}
because $J_3$ is usually coupled to $J_1$ and $J_2$. 
While our approach could in principle be generalized to include highly non-Gaussian initial conditions, 
we have assumed Gaussian statistics throughout most of this paper since the current cosmological data 
indicates that the primeval fluctuations were very close to Gaussian.
As an illustration, we have computed the non-Gaussian peak bias induced by a small, local primordial 
trispectrum and shown that our findings are consistent with a ``peak-background split'' expectation.

We have not included the effect of gravitational motions from the initial (Lagrangian) to 
final (Eulerian) positions. This complication was addressed in \cite{desjacques/crocce/etal:2010} using
the Zeldovich approximation solely. Note that, at third order in perturbation theory, a term identical
to $J_3(\vx)$ arises 
\cite{chan/scoccimarro/sheth:2012,biagetti/desjacques/etal:2014,saito/baldauf/etal:2014}, except that 
$\delta_R$ should be replaced by the gravitational potential $\phi$. 
In practice, the integrated perturbation theory (iPT) \citep{matsubara:2011,matsubara:2014} would 
provide a more systematic way of computing these corrections at a given order in Lagrangian PT.
All this is left to future work.

\section*{Acknowledgment}

V.D. would like to thank Dmitry Pogosyan for interesting discussions, and acknowledges support by the 
Swiss National Science Foundation. M.M. thanks Ravi Sheth for many insightful comments on multivariate bias expansions.

\appendix

\section{Peak-Background split approach} 
\label{sec:PBSa}

In this Appendix, we will show that Eq.(\ref{eq:DB1}) agrees with the result obtained from a peak-background 
split (PBS).
In the PBS, the non-Gaussian bias induced by a primordial trispectrum is proportional to the derivative of 
the non-Gaussian halo mass function w.r.t. the skewness of the density field
\cite{desjacques/jeong/schmidt:2011,smith/ferraro/loverde:2012}. In practice, we will take advantage of the 
smallness of the PNG to model the non-Gaussian mass function with a Gram-Charlier expansion.
 
\subsection{Gram-Charlier expansion at third order}
\label{subsec:GCexp3}

The derivation of the Gram-Charlier series expansion in terms of rotational invariants of a random field is
presented in \cite{pogosyan/gay/pichon:2009,gay/pichon/pogosyan:2012} (see, in particular, Eq.(24) of the
second reference). 
For generic non-Gaussian initial conditions and for the invariants $\vw=(\nu,J_1,3\eta^2,5J_2,J_3)$ relevant 
here, the non-Gaussian contribution $\delta P_\text{NG}$ to the joint PDF 
$P_\text{NG}(\vw)=\mathcal{N}(\vw)+\delta P_\text{NG}(\vw)$ is
\begin{align}
\label{eq:dPNG}
\delta &P_\text{NG}(\nu,J_1,3\eta^2,5J_2,J_3) = \mathcal{N}(\nu,J_1,3\eta^2,5J_2,J_3) \\
&\times\Bigg[\frac{1}{6}\left\langle \nu^3\right\rangle_\text{\tiny GC} H_{30}(\nu,J_1)
+\frac{1}{6}\left\langle J_1^3 \right\rangle_\text{\tiny GC} H_{03}(\nu,J_1) \nonumber \\
&\quad
+ \frac{1}{2}\left\langle J_1\nu^2 \right\rangle_\text{\tiny GC} H_{21}(\nu,J_1)
+ \frac{1}{2}\left\langle \nu J_1^2 \right\rangle_\text{\tiny GC} H_{12}(\nu,J_1) \nonumber \\
&\quad 
- \left\langle \nu \eta^2 \right\rangle_\text{\tiny GC} H_{10}(\nu,J_1) 
L^{(1/2)}_1\!\left(\frac{3}{2}\eta^2\right) \nonumber \\
&\quad 
- \left\langle \nu J_2 \right\rangle_\text{\tiny GC} H_{10}(\nu,J_1) 
L^{(3/2)}_1\!\left(\frac{5}{2}J_2\right) \nonumber \\
&\quad
- \left\langle J_1 \eta^2\right\rangle_\text{\tiny GC} H_{01}(\nu,J_1) 
L^{(1/2)}_1\!\left(\frac{3}{2}\eta^2\right)  \nonumber \\
&\quad
- \left\langle J_1 J_2 \right\rangle_\text{\tiny GC} H_{01}(\nu,J_1) 
L^{(3/2)}_1\!\left(\frac{5}{2}J_2\right)
+ \frac{25}{21}\left\langle J_3 \right\rangle_\text{\tiny GC} J_3 \Bigg] \;, \nonumber
\end{align}
at leading order. 
The ensemble average $\big\langle X\big\rangle_\text{\tiny CG}$ is a shorthand for the moments
\begin{align}
\label{eq:GC1}
\big\langle &\nu^i J_1^j \eta^{2q} J_2^l J_3^m\big\rangle_\text{\tiny GC} = 
\frac{(-1)^qq!(-1)^ll!}{(3/2)^q(5/2)^l}\,\frac{1}{\bnpk} \\
&\times
\left\langle \npk(\vw)\,H_{ij}(\nu,J_1) L^{(1/2)}_q\!\left(\frac{3}{2}\eta^2\right) 
L^{(3/2)}_l\!\left(\frac{5}{2}J_2\right) J_3^m\right\rangle \nonumber \;,
\end{align}
where we have assumed $m\leq 1$. The normalization coefficient can generally be extracted by projection 
as explained in \cite{pogosyan/gay/pichon:2009}. 
Furthermore, it is straightforward to show that, up to third order, the Gram-Charlier expansion is equal 
to the Edgeworth series of the same order, i.e. 
$\big\langle y^3 \big\rangle_\text{\tiny GC}\equiv \big\langle y^3\big\rangle_c$.

\subsection{Moments induced by a long mode}
\label{subsec:CorrelPBS}

In order to apply the peak-background split argument, we must think of the expectation values
$\bigl\langle y^3\bigr\rangle_c$ as the third order moments induced by a long-wavelength 
perturbation $\phi_\text{\tiny L}(\vk_\text{\tiny L})$ in the gravitational potential 
\cite{desjacques/jeong/schmidt:2011}. 
We will designate them as $\left\langle X_2\right\rangle_\text{\tiny L}$ (as it is obvious that
they correspond to the fields defined at $\vx_2$ in Eq.(\ref{eq:schematic})).

To evaluate these moments, we substitute the Fourier representation of $X_2$ and take advantage of the fact 
that, for cubic-order local PNG, the 3-point function of (small-scale) density fluctuations induced by a 
single perturbation $\phi_\text{\tiny L}(\vk_\text{\tiny L})$ is \cite{desjacques/jeong/schmidt:2011}
\begin{align}
\big\langle
\delta_R(\mb{k}_1)\delta_R(\mb{k}_2)&\delta_R(\mb{k}_3)
\big\rangle_\text{\tiny L}
= 6 \gnl \phi_\text{\tiny L}(\vk_\text{\tiny L}){\cal M}_R(k_1) {\cal M}_R(k_2) \nonumber \\ 
&\qquad \times {\cal M}_R(k_3)\Bigl[ P_\phi(k_1) P_\phi(k_2) + \mbox {2 cyc.} \Bigr] 
\nonumber \\
& \qquad \times (2\pi)^3\delta_D\!(\vk_1+\vk_2+\vk_3) \;.
\end{align}
After simplification, the result is of the form
\begin{align}
\label{eq:momentphiL}
\big\langle X_2\big\rangle_\text{\tiny L} &= 
6\gnl  \phi_\text{\tiny L}(\vk_\text{\tiny L}) g(\sigma_0,\sigma_1,\sigma_2) \\
&\qquad \times \frac{1}{\sigma_0^4}\iint \frac{d^3k_1}{(2 \pi)^3}\frac{d^3k_2}{(2 \pi)^3}\, 
\mathcal{M}_R(k_1) \mathcal{M}_R(k_2) \nonumber \\
&\qquad \times \mathcal{M}_R(k_{12}) {\cal X}_2(\vk_1,\vk_2)\Big[ P_\phi(k_1) P_\phi(k_2)+\mbox{2 cyc.}\Big] 
\nonumber \\
&= 3 \gnl \phi_\text{\tiny L}(\vk_\text{\tiny L}) S_3^{(X_2)}\!(M) g(\sigma_0,\sigma_1,\sigma_2)
\nonumber \;.
\end{align}
Here again, we took advantage of the fact that the cyclic sum of $P_\phi(k_1)P_\phi(k_2)$ is invariant under
any permutation of $k_1$, $k_2$ and $k_{12}$.

It is instructive to consider the effect of a long-wavelength perturbation on the determinant of $\zeta_{ij}$,
which appears in the expression of $J_3$, Eq.(\ref{eq:J3}). In this case, we find
\begin{widetext}
\begin{align}
\left\langle\mbox{det}(\zeta_{ij})\right\rangle_\text{\tiny L} = & - \frac{1}{\sigma^3_2} 
\int\!\!\frac{d^3k}{(2 \pi)^3}\int\!\!\frac{d^3k'}{(2 \pi)^3}\int\!\!\frac{d^3k"}{(2 \pi)^3}\, 
\epsilon_{ijk}\, k_1 k_i k'_2 k'_j k^{_{"}}_3 k^{_{"}}_k 
\big\langle \delta_R(\mb{k}) \delta_R(\mb{k}')\delta_R(\mb{k}")\big\rangle_\text{\tiny L} 
e^{i\left(\mb{k}+\mb{k}'+\mb{k}"\right)\cdot\mb{x}} \nonumber \\
=& \frac{6 \gnl}{\sigma_2^3}\phi_\text{\tiny L}(\vk_\text{\tiny L}) 
\int\!\!\frac{d^3k}{(2 \pi)^3} \int\!\!\frac{d^3k'}{(2 \pi)^3} 
\mathcal{M}_R(k)\mathcal{M}_R(k')\mathcal{M}_R(|\mb{k}+\mb{k}'|) \nonumber \\
&\qquad \times \Bigl[P_\phi(k) P_\phi(k')+ P_\phi(k') P_\phi(|\vk+\vk'|)+P_\phi(|\vk+\vk'|)P_\phi(k)\Bigr]
\nonumber \\
&\qquad
\times k_1 k'_2 (-\mb{k}-\mb{k}')_3 \,\Bigl[\mb{k}\wedge \mb{k}'\cdot\left(-\mb{k}-\mb{k}'\right)\Bigr] \;.
\label{eq:geomDet}
\end{align}
\end{widetext}
To obtain the last equality, we have used the fact that 
$\epsilon_{ijk}k_i k'_j k^{_{"}}_k = \mb{k}\wedge\mb{k}'\cdot\mb{k}"$. 
This last product is just the volume of the 3-dimensional parallelepiped generated by the wavemodes $\mb{k}$, 
$\mb{k}'$, and $\mb{k}"$. However, momentum conservation imposes the condition  $\mb{k}" = - (\mb{k}+\mb{k}')$ 
or, equivalently, that these wavemodes be in the same plane. Therefore, the volume is trivially zero, so that 
$\left\langle\mbox{det}(\zeta_{ij})\right\rangle_\text{\tiny L}=0$.

\subsection{Local fluctuations in peak abundance}
\label{subsec:localdeltapk}

At this point, we substitute Eq.(\ref{eq:momentphiL}) into Eq.(\ref{eq:dPNG}) and recast the orthogonal 
polynomials in terms of the bias parameters. 
For instance, the term proportional to $\big\langle J_1^3\big\rangle_c$ becomes
\begin{align}
\frac{1}{6}\big\langle J_1^3\big\rangle_c H_{03}(\nu,J_1) &= \frac{1}{\sqrt{6}}
\big\langle J_1^3\big\rangle_c \tilde{H}_{03}(\nu,J_1) \\
&= 3 \gnl \phi_\text{\tiny L}(\vk_\text{\tiny L})\sigma_0^4 S_3^{(J_1^3)}
\frac{\tilde{H}_{03}(\nu,J_1)}{\sigma_2^3\sqrt{6}}  \nonumber \;,
\end{align}
since we have $g(\sigma_0,\sigma_1,\sigma_2)=\sigma_0^4/\sigma_2^3$ for this particular example. 

To obtain the response $\bar{n}_{\rm pk}^\text{\tiny L}$ of the peak number density to the skewness 
induced by a local cubic non-Gaussianity, we now multiply this last expression by 
$\cpk(\vy)=\cpk(\vw)$ and integrate over the variables $\vw$. 
Orthogonal polynomials are replaced by the peak bias factors $c_{ijqlm}$ according to Eqs.~(\ref{eq:cijqlm})
and (\ref{eq:ctildec}).
After some algebra, the correction $\delta\bnpk^\text{\tiny L}(\vk_\text{\tiny L})$ to the peak abundance 
at first order in the long-wavelength perturbation $\phi_\text{\tiny L}(\vk_\text{\tiny L})$ is
\begin{align}
\label{eq:deltaPNG}
\delta\bar{n}_{\rm pk}&^\text{\tiny L}(\vk_\text{\tiny L})  = 
3 \gnl\phi_\text{\tiny L}(\vk_\text{\tiny L})\bnpk \sigma_0^4 \\
&\times 
\Bigg\{\frac{1}{6}c_{30000} S_3^{(\nu^3)} + \frac{1}{2} c_{21000} S_3^{(J_1\nu^2)} 
+\frac{1}{2} c_{12000} S_3^{(\nu J_1^2)} \nonumber \\
&\quad +\frac{1}{6} c_{03000} S_3^{(J_1^3)} + c_{10100} S_3^{(\nu\eta^2)} 
+ c_{01100} S_3^{(J_1\eta^2)} \nonumber \\
&\quad + c_{10010} S_3^{(\nu J_2)} + c_{01010} S_3^{(J_1J_2)}
+\frac{5}{\sqrt{7}}c_{00001}S_3^{(J_3)} \Bigg\} 
\nonumber \\
&\equiv 3\gnl \phi_\text{\tiny L}(\vk_\text{\tiny L})\, \delta\bar{n}_{\rm pk}^{\fnl=1} 
\nonumber \;,
\end{align}
where $\delta\bar{n}_{\rm pk}^{\fnl=1}$ is the non-Gaussian correction to the halo mass function in the 
presence of local quadratic non-Gaussianity with $\fnl=1$. 

The last equality follows from the definition of the average peak number density, Eq.(\ref{eq:bnpk}). 
Namely, the non-Gaussian correction generally is
\begin{align}
\label{eq:dnpkNG}
\delta\bnpk^\text{NG} &=\int\!\!d\vw\,\cpk(\vw)\, \delta P_\text{NG}(\vw) \\
&= \int\!\!d\vw\,\cpk(\vw)\, P_\text{G}(\vw) \nonumber \\
&\qquad \times \bigg[\frac{1}{6}\big\langle\nu^3\big\rangle_c H_{30}(\nu,J_1) +\dots + \frac{25}{21}
\big\langle J_3\big\rangle_c J_3\bigg] \nonumber \;,
\end{align}
which follows from the substitution of Eq.(\ref{eq:dPNG}). Specializing to local, quadratic PNG for which
the bispectrum is 
\begin{equation}
B_\phi(k_1,k_2,k_3)=2\fnl\Bigl[P_\phi(k_1) P_\phi(k_2)+ \mbox{2 cyc.}\Bigr] \;,
\end{equation}
the third-order moments in Eq.(\ref{eq:deltaPNG}) are given by
\begin{align}
\big\langle X_2\big\rangle_c &= \frac{g(\sigma_0,\sigma_1,\sigma_2)}{\sigma_0^4}
\iint\!\!\frac{d^3k_1}{(2\pi)^3}\frac{d^3k_2}{(2\pi)^3}\,\mathcal{X}_2(\vk_1,\vk_2)
\nonumber \\
&\qquad \times B_R(k_1,k_2,k_{12}) \nonumber \\
&= 2\fnl\frac{g(\sigma_0,\sigma_1,\sigma_2)}{\sigma_0^4}
\iint\!\!\frac{d^3k_1}{(2\pi)^3}\frac{d^3k_2}{(2\pi)^3}\,\mathcal{X}_2(\vk_1,\vk_2)
\nonumber \\
&\qquad \times \Big. {\cal M}(k_1){\cal M}(k_2){\cal M}(k_{12}) \nonumber \\
&\qquad \times \Big[P_\phi(k_1) P_\phi(k_2)+\mbox{2 cyc.}\Big]
\nonumber \\
&= \fnl\, g(\sigma_0,\sigma_1,\sigma_2)\, S_3^{(X_2)} \;,
\end{align}
where $B_R$ is the bispectrum of the smoothed, linear density field.
Performing the integral in Eq.(\ref{eq:dnpkNG}) and using the definition of the bias coefficients $c_{ijqlm}$,
we arrive at
\begin{align}
\delta\bnpk^{f_\text{NL}} &= \fnl\bnpk \sigma_0^4\bigg[\frac{1}{6}c_{30000} S_3^{(\nu^3)}
+\dots + \frac{5}{\sqrt{7}} c_{00001} S_3^{(J_3)}\bigg]\;,
\end{align}
which is precisely the right-hand side of Eq.(\ref{eq:deltaPNG}) provided that 
$\fnl\equiv 3\gnl\phi_\text{\tiny L}(\vk_\text{\tiny L})$. 
As explained in \cite{desjacques/jeong/schmidt:2011}, this reflects the fact that, in the cubic model, a 
long-wavelength mode locally generates a 3-point function with amplitude proportional to 
$\phi_\text{\tiny L}$.

\bibliographystyle{prsty}
\bibliography{references}

\label{lastpage}

\end{document}